\documentclass[11pt,prd,superscriptaddress,nofootinbib]{revtex4}
\usepackage[english]{babel}
\usepackage{graphicx}
\usepackage{mathtext}
\usepackage{indentfirst}
\usepackage{epsfig,amsmath,amsfonts}

\newcommand{\beq}[1]{\begin{eqnarray}\label{#1}}
\newcommand\eeq {\end{eqnarray}}
\newcommand\bqa {\begin{eqnarray}}
\newcommand\eqa {\end{eqnarray}}
\newcommand\pr {\partial}

\newcommand{\eq}[1]{(\ref{#1})}
\newcommand{\bear}{\begin{array}}
\newcommand{\enar}{\end{array}}

\newcommand{\cH}{{\cal H}}
\newcommand{\K}{\mathcal{K}}
\newcommand{\F}{\mathbb{F}}
\newcommand{\A}{\mathbb{A}}
\newcommand{\Z}{\mathbb{Z}}

\newcommand{\N}{\mathbb{N}}

\begin{document}

\centerline{\Large\bf IR divergences and kinetic equation in de Sitter space.}
\centerline{\Large \bf (Poincare patch; Principal series)}

\vspace{5mm}

\centerline{\bf E.T.Akhmedov}

\vspace{5mm}

\centerline{B.Cheremushkinskaya, 25, ITEP, Moscow, 117218, Russia}

\vspace{3mm}

\centerline{and}

\vspace{3mm}

\centerline{MIPT, Dolgoprudny, Moscow Region, Russia}

\vspace{5mm}

\centerline{\bf Abstract}

\vspace{3mm}

We explicitly show that the one loop IR correction to the two--point function in de Sitter space scalar QFT does not reduce just to the mass renormalization. The proper interpretation of the loop corrections is via particle creation revealing itself through the generation of the quantum averages $\langle a^+_p a_p\rangle$, $\langle a_p a_{-p}\rangle$ and $\langle a^+_p a^+_{-p}\rangle$, which slowly change in time. We show that this observation in particular means that loop corrections to correlation functions in de Sitter space can not be obtained via analytical continuation of those calculated on the sphere.

We find harmonics for which the particle number $\langle a^+_p a_p\rangle$ dominates over the anomalous expectation values $\langle a_p a_{-p}\rangle$ and $\langle a^+_p a^+_{-p}\rangle$. For these harmonics the Dyson--Schwinger equation reduces in the IR limit to the kinetic equation. We solve the latter equation, which allows us to sum up all loop leading IR contributions to the Wightman function. We perform the calculation for the principal series real scalar fields both in expanding and contracting Poincare patches.

\vspace{5mm}


\section{Introduction}

There are large infrared (IR) loop contributions to the correlation functions even due to massive fields in de Sitter (dS) space. The question is whether after the summation over all leading IR loop contributions the correlation functions will be finite or will grow unboundedly. If the correlation functions do grow in the IR limit, then this is the sign that back--reaction of the quantum fluctuations on the background dS geometry is not negligible. The literature on this subject is vast, see e.g. \cite{Mottola:1984ar}--\cite{Akhmedov:2008pu}.

In general to sum the loop contribution one has to solve the Dyson--Schwinger equation (DSE).
However, the interest is in the solution of this equation in the extreme IR limit, i.e. in the sum of the leading IR corrections. Colloquially speaking in the latter limit quantum coherence is lost
and the legs in the loops sit on mass--shell. The situation is analogous to the one discussed in a similar context e.g. in \cite{Tsamis:2005hd}--\cite{vanderMeulen:2007ah}. As the result, known in condensed matter physics, the DSE for non--stationary diagrammatic technic reduces to the Boltzmann equation (see e.g. \cite{LL}, \cite{Kamenev}). In other words the classical Boltzmann kinetic equation (KE) allows to sum up the leading IR corrections. We concisely review the relation between DSE and KE in the Appendix.

In this note we derive the Boltzmann equation in the Poncare patch (PP) of dS space. For the earlier papers on the KE in curved spaces in general and in dS space in particular see e.g. \cite{Calzetta:1986ey},\cite{Kitamoto:2010si}.  Unlike the equations of those papers our equation does not really describe the dynamics of the occupation numbers of particles in dS space, because in its derivation we use the exact harmonics on dS background rather than the plain waves. It describes the dynamics of the occupation numbers of the waves with wavelength comparable to the horizon size. We would like to study the physics due to long wavelength fluctuations because IR effects are dominated by such objects. Still we call them as particles throughout the paper.

Our eventual goal is to see whether the back--reaction from the quantum fluctuations in dS space is negligible or not. But for the beginning we neglect back--reaction and consider self--interacting massive QFT on the fixed dS background. We restrict our attention to the behavior of the solution of the KE in the extreme IR limit.

The consideration of the exact harmonics in the background fields brings several complications.
The main one is generic for the curved space--times and is due to that even for the fundamental questions in curved space--times it is necessary to specify boundary conditions, so to say. This is because there is no any preferable reference frame in a general curved space--time.

\begin{figure}[t]
\begin{center} \includegraphics[width=300pt]{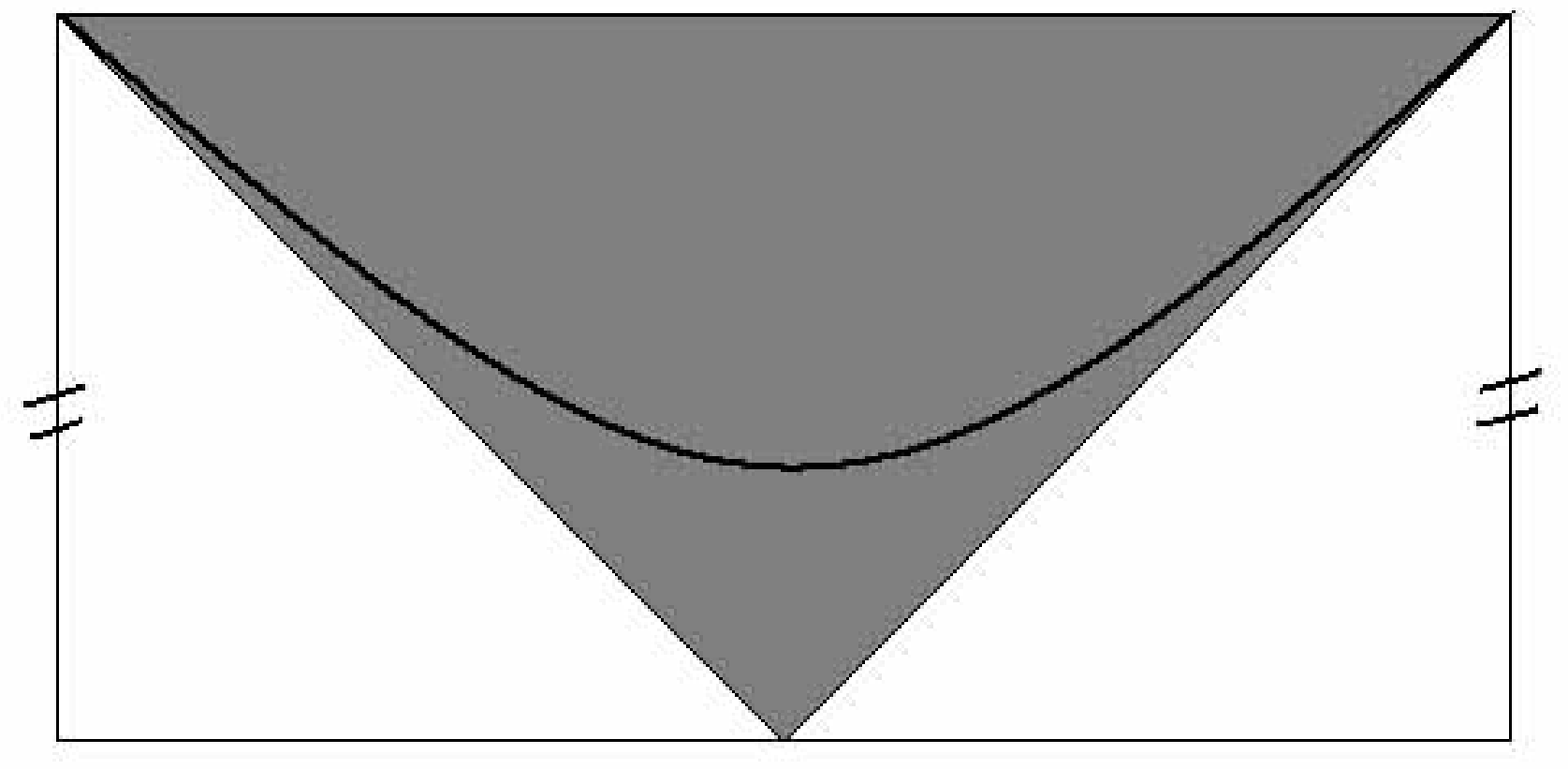}
\caption{}
\end{center}
\end{figure}

\begin{figure}[t]
\begin{center} \includegraphics[width=300pt]{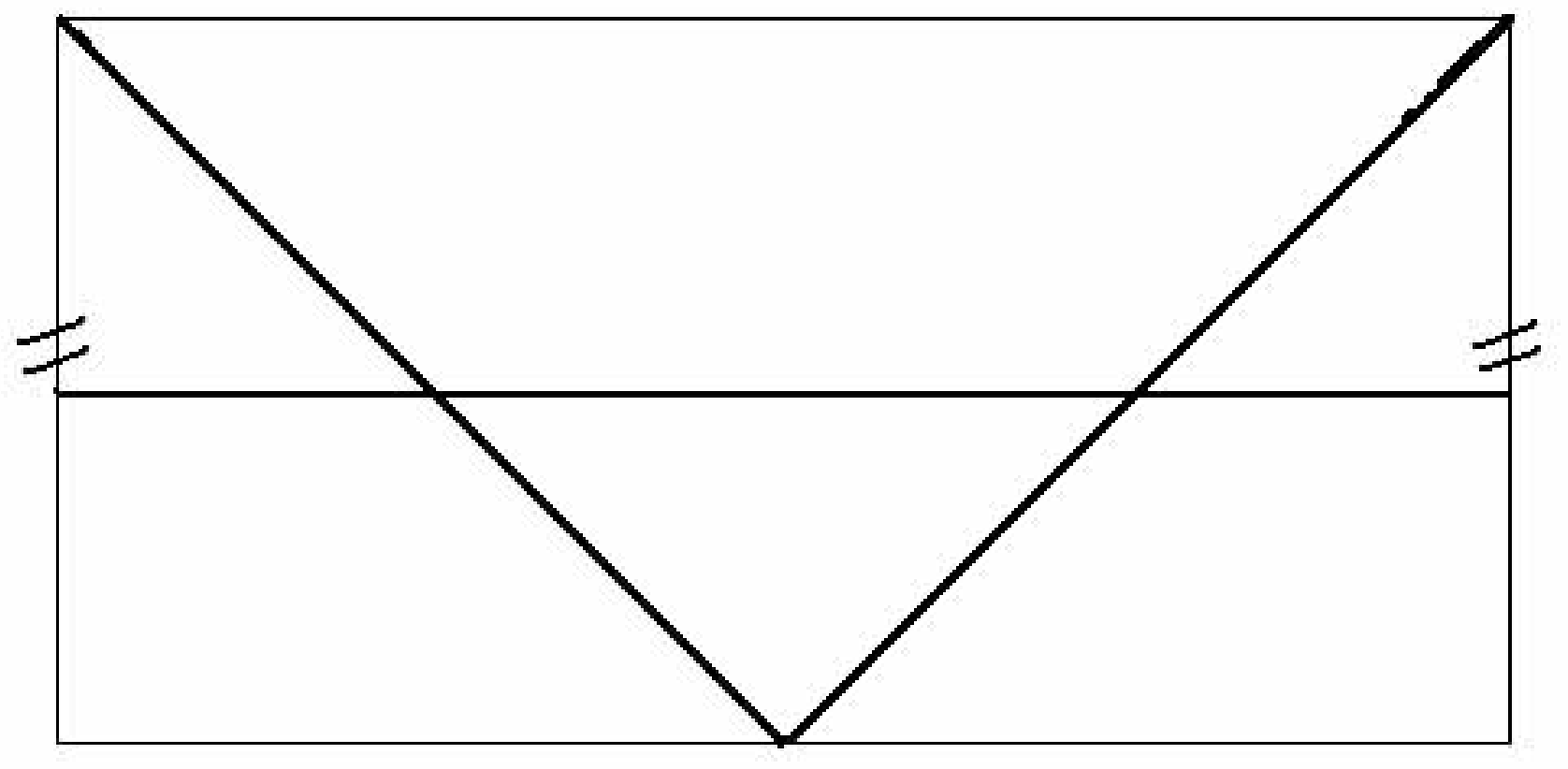}
\caption{}
\end{center}
\end{figure}

In dS space this complication reveals itself through the crucial difference between Cauchy problems for the KE as defined in global dS and in PP. The main difference is due to the different geometry of the Cauchy surfaces in these two situations. The PP of dS space is the grey region in the rectangle shown on the figure 1 --- Penrose diagram
of the dS space. The left and right sides of the rectangle are glued to each other. The Cauchy surfaces in the PP are depicted as the lines close to the hyperbolas shown in this
figure. As we go to the past infinity the Cauchy surfaces degenerate to the boundary of the PP. The Bunch--Davies (BD) vacuum means that there are no positive energy states on the boundary of the PP. Complementary to the grey part of the diagram is the contracting PP.

The Cauchy surfaces in global dS space are depicted by the lines of the type shown on the figure 2. E.g. the Euclidian vacuum means, so to say, that there are no positive energy states on
such a surface going nearby the neck of dS space. In such a situation we have a different state at the boundary of the PP.
Thus, even despite that the Wightman function for BD vacuum state in PP coincides with the one for Euclidian vacuum in global dS, the QFT dynamics in both situations can be quite different. In particular,
global dS space contains as well the contracting PP. We point out that the behavior of solutions of the KE in the contracting patch is quite different from the expanding one. Furthermore, from the cosmological point of view one needs to study global dS because it sets up initial conditions for the PP --- for the inflation. But this issue demands a separate study and will not be addressed in the present paper. Here we restrict our attention to the expanding and contracting PP of dS space.

Another complication is peculiar for homogeneous time--dependent backgrounds with horizons such as dS space. It comes from the fact that unlike Minkowski and AdS space there is no unique choice of harmonic basis which diagonalizes free Hamiltonian once and forever. To address this complication we do not specify harmonics in the general formulas. We attempt to use formulas which have a generic application in dS space or even for more general FRW space--times with flat spacial sections. At the end we of cause study physics due to different specific choices of harmonics.

Yet another complication is as follows. The regular KE for particles describes the change in time of
the particle density $n_p = \langle a^+_{\vec{p}} \, a_{\vec{p}} \rangle$. At the same time one assumes that the anomalous quantum averages $\kappa_p = \langle a_{\vec{p}} \, a_{-\vec{p}}\rangle$ and $\kappa^*_p = \langle a^+_{\vec{p}} \, a^+_{-\vec{p}}\rangle$ are vanishingly small. In background fields in general this is not the case any more. Only if one finds the appropriate equilibrium state he can neglect $\kappa_p$. In the paper \cite{Polyakov:2009nq} the KE of the type interesting for us was derived. However in that paper the analysis of the presence of the anomalous quantum averages $\kappa_p$ and $\kappa^*_p$ was not done. As the result the wrong choice of the harmonics was made there.

Furthermore, waggly speaking KE is valid if its solution $n_p$ is slow function of time in comparison with the corresponding harmonics. The consideration of the long wavelength fluctuations poses the question about the existence of such a separation of scales in KE. However, we check whether $n_p$ is slow enough or not for every explicit solution of the equation, which we find.

Finally, in this paper we discuss the scalar fields from the principal series, which have masses $m>3/2$ in units of the dS curvature. The dynamics of the light fields, $m<3/2$, i.e. from the complementary series, can be quite different. It goes without saying about the massless fields.

\section{Specification of harmonics}

The goal of this paper is to derive kinetic equation (KE) for the exact harmonics in de Sitter
(dS) space. To do that we have to specify which harmonics one should use. As we explain in the next section the proper KE can be derived only in such circumstances when one can neglect $\langle a_p \, a_{-p}\rangle$ in comparison with $\langle a^+_p \, a_p \rangle$, where $a$ and $a^+$ are annihilation and creation operators. Below we will be able to specify such harmonics for which $\langle a_p \, a_{-p}\rangle$ is negligible in comparison with $\langle a^+_p \, a_p\rangle$ at the low physical momenta $p\eta\to 0$, where $\eta$ is the conformal time in the dS space\footnote{The extreme IR limit for the physical momentum corresponds to the future infinity for the fixed comoving momentum in the expanding Poincare patch (PP) of dS space.}.

The way we will do that is as follows.
The Keldysh propagator carries statistical information about the theory and the state in question (see the discussion below). The calculation of the one loop contribution to the Keldysh propagator will allow us to estimate the behavior of $\langle a^+_p a_p\rangle$ and $\langle a_p a_{-p}\rangle$. For some harmonics $\langle a^+_p \, a_p \rangle$ will be of the same order as $\langle a_p \, a_{-p}\rangle$, while for the others $\langle a_p \, a_{-p}\rangle$ will be suppressed in the IR limit.

We would like to consider scalar field theory in the background space--time with the metric:

\bqa
ds^2 = a^2(\eta)\, \left[d\eta^2 - d\vec{x}^2\right].
\eqa
In this paper we always consider $(+,-,-,-)$ signature. Although all our formulas can be straightforwardly generalized to arbitrary dimensions we restrict ourselves to 4D.

The choice $a(\eta) = 1/\eta$, $0<\eta = e^{-t}<+\infty$ corresponds to the PP of dS space. We put the Hubble constant to one $H=1$. Past infinity of the PP corresponds to $\eta \equiv e^{-t}\to + \infty$, while the future infinity is at $\eta \equiv e^{-t} \to 0$. We prefer the definition $\eta = e^{-t}>0$ instead of the more standard one $\eta = - e^{-t}<0$ to avoid dealing with non--integer powers of the negative quantity. The contracting PP corresponds as well to $a(\eta) = 1/\eta$, but now $\eta = e^t$ and past infinity corresponds to $\eta = 0$, while future infinity --- to $\eta = + \infty$. So in contracting PP conformal time flows in the reverse direction with respect to (wrt) the expanding PP.
Below we always talk about either expanding or contracting PP of dS space, but prefer to keep generic $a(\eta)$ in those formulas which may be applied to more general space--times.

We consider the theory of the real massive scalar particle with the cubic self--interaction:

\bqa\label{action}
S \equiv \int d^3x \int_{\infty}^0 d\eta \, \sqrt{|g|}\,{\cal L}[\phi] = \int d^3x\, \int_{\infty}^0 d\eta \, a^4(\eta)\, \left[\frac{1}{2\, a^2(\eta)} \,\left(\pr_\mu \phi\right)^2 - \frac{m^2}{2} \, \phi^2 - \frac{\lambda}{3}\, \phi^3\right].
\eqa
Although the cubic potential has a run away instability we have chosen it to simplify
all our formulas. A short comment on how the instability of the cubic theory reveals itself in the KE can be found in the first part of the Appendix.
Conceptually all our calculations are valid as well for the theory with $\phi^4$
self--interaction term. In fact, it is not difficult to write the answer for the latter theory
once the answer for the cubic self--interaction is known.

\subsection{Tree--level two--point function}

To set up the notations we have to start with the tree--level two--point function.
The expansion of the free field in terms of the normalized harmonics is

\bqa
\phi(\eta, \vec{x}) = \int d^3k \left[a_k \, g_k(\eta) \, e^{-i\, \vec{k}\, \vec{x}} + a^+_k \, g^*_k(\eta) \, e^{i\, \vec{k}\, \vec{x}}\right],
\eqa
where $\phi_k (\eta, x) = g_k(\eta)\, e^{-i\, \vec{k}\, \vec{x}}$ are some solutions of the Klein--Gordon equation,

\bqa
\left[a^{-4}\, \pr_\eta \,a^2\,\pr_\eta - \frac{\Delta}{a^2} + m^2\right]\, \phi(\eta,x) = 0,
\eqa
in the metric under consideration.
Concretely $g_k(\eta) = \eta^{3/2}\, h(k\eta)/\sqrt{2}$, where $h(k\eta)$ is some properly normalized solution of the Bessel equation, $a(\eta) = 1/\eta$.

From the energy momentum tensor $T_{\mu\nu} = \pr_\mu \phi \pr_\nu \phi - g_{\mu\nu} \, {\cal L}[\phi]$, we find the Hamiltonian: $H(\eta) = a^2(\eta) \, \int d^3x \, T_{00}(\eta).$ The free Hamiltonian looks as\footnote{After the normal ordering of the free Hamiltonian one has to absorb the infinite vacuum energy into the redefinition of the cosmological constant. Such a quantity is time independent if expressed in terms of the physical momentum $p/a(\eta)$ and attributed to the Hamiltonian conjugate to the time $t$ rather than to the conformal time $\eta = e^{-t}$.}:

\bqa\label{freeham}
H_0(\eta) = \int d^3 k \, \left[a^+_k \, a_k \, A_k(\eta) + a_k \, a_{-k}\, B_k(\eta) + h.c.\right], \nonumber \\
A_k(\eta) = \frac{a^2(\eta)}{2} \, \left\{\left|\frac{dg_k}{d\eta}\right|^2 + \left[k^2 + a^2(\eta)\, m^2\right]\, \left|g_k\right|^2\right\}, \nonumber \\
B_k(\eta) = \frac{a^2(\eta)}{2}\, \left\{\left(\frac{dg_k}{d\eta}\right)^2 + \left[k^2 + a^2(\eta)\, m^2\right]\, g_k^2\right\}.
\eqa
The harmonics, which simultaneously solve the Klein--Gordon equation and $B_k=0$ can be found only in those regimes, when the background field is switched off and, hence, $A_k$ and $B_k$ are time independent. Then the harmonics just coincide with the plain--waves. In general the solutions of the equation $B_k = 0$ and of the Klein--Gordon one do not coincide, because $B_k$ is time dependent. Hence, the free Hamiltonian is not diagonal. Furthermore, one can not choose the solutions of the equation $B_k=0$ as $g_k$, because then the corresponding $a_k$ and $a_k^+$ do not obey the Heisenberg algebra.

We do not specify harmonics in the formulas which have general application, but below we are going to consider particular cases. Let us make a few comments about the standard choice of the Bunch--Davies (BD) harmonics in the PP \cite{Bunch:1978yq}:

\bqa\label{BD}
g_k(\eta) \equiv \frac{\sqrt{\pi} \, \eta^{\frac32}\, e^{- \frac{\pi \, \mu}{2}}}{2}\, \cH^{(1)}_{i\,\mu}(k\, \eta), \quad k = \left|\vec{k}\right|, \quad \mu = \sqrt{m^2 - \frac94},
\eqa
where $\cH_{i\mu}^{(1)}(x)$ is the Hankel function. We choose $\cH_{i\mu}^{(1)}(x)$ as the positive energy harmonics instead of $\cH_{i\mu}^{(2)}(x)$ because of the reverse order of the time flow: $\infty \to \eta \to 0$. It is not hard to see that the harmonics in question diagonalize the Hamiltonian at the past infinity $\eta\to \infty$, because in this limit $g_k(\eta) \to \frac{\eta}{\sqrt{k}} \, e^{i\,k\, \eta}$ and $B_k\to 0$, while $A_k \to k$.  In the future infinity $\eta \to 0$ there is a solution of the Klein--Gordon equation, which only approximately solves $B_k=0$ for the very massive fields $m\gg 3/2$.

The Keldysh propagator is defined as\footnote{Note that here we use a slightly different definition of the diagrammatic rules than in the Appendix. The difference is in the Keldysh rotation --- instead of the one used in the Appendix here we use $\phi_{cl} = (\phi_+ + \phi_-)/2$, $\phi_q = \phi_+ - \phi_-$. The rules in the main body of the text agree e.g. with \cite{Leblond}, while those used in the Appendix agree with \cite{Kamenev}. That is because here we would like to compare our formulas to \cite{Leblond}, while the formulas in the Appendix should be compared to \cite{Kamenev}.}
$D^K(\eta_1,\eta_2,\left|\vec{x} - \vec{y}\right|) = \frac12 \, \left\langle \left\{\phi(\eta_1, \vec{x}),\, \phi(\eta_2, \vec{y})\right\} \right\rangle.$
Due to the spacial homogeneity of dS space we find it more convenient to make the Fourier transform along the spacial directions --- $D^K_p(\eta_1, \eta_2) \equiv \int d^3r \, D^{K}(\eta_1, \eta_2, \vec{r}) \, e^{-i\, \vec{p}\, \vec{r}}$. Furthermore, to simplify the formulas we define $d^K(p\eta_1,\, p\eta_2) \equiv \left(\eta_1\,\eta_2\right)^{-\frac32}\, D^K_p(\eta_1, \eta_2)$. At tree--level this quantity is equal to:

\bqa
d^K_0(p\eta_1, p\eta_2) = \frac{1}{2} \, \left[h(p\eta_1)\, h^*(p\eta_2) + h^*(p\eta_1) \, h(p\eta_2)\right]\, \left(1 + 2\, \left\langle a^+_p \, a_p\right\rangle\right) + \nonumber \\ + h(p\eta_1)\, h(p\eta_2)\, \left\langle a_p \, a_{-p}\right\rangle + h^*(p\eta_1)\, h^*(p\eta_2)\, \left\langle a^+_p \, a^+_{-p}\right\rangle,
\eqa
if we average wrt an arbitrary spatially homogeneous state. If the average is taken wrt the vacuum state $a_p \, |\rangle = 0$, then $\left\langle a_p \, a_{-p}\right\rangle = \left\langle a^+_p \, a^+_{-p}\right\rangle = 0$, and $\left\langle a^+_p \, a_p\right\rangle = 0$.
Different choices of the harmonics $h(p\eta)$
correspond to the different choices of the vacuum state.
Using Gradshtein and Rizhik eq. 6.672.2, one can show that for the arbitrary choice of the solution of the Bessel equation $h(p\eta)$ the Keldysh propagator away from the singularity is equal to:

\bqa
D^K(\eta_1,\eta_2,\left|\vec{x} - \vec{y}\right|) = C_1 \,\left(z^2 - 1\right)^{-\frac12}\, P^1_{-\frac12 + i\, \mu}(z) + C_2 \, \left(z^2 - 1\right)^{-\frac12}\, Q^1_{-\frac12 + i\, \mu}(z),
\eqa
where $P^1_\nu$ and $Q^1_\nu$ are associated Legendre functions; $z = 1 + \frac{(\eta_1 - \eta_2)^2 - |\vec{x} - \vec{y}|^2}{2\eta_1 \, \eta_2}$ is the hyperbolic distance between the two points in question on dS space; $C_{1,2}$ are some complex constants whose values depend on the particular choice of the Harmonics. E.g. for the BD harmonics \eq{BD} $C_2=0$ and the two--point function coincides with the one following from the analytical continuation from the sphere and having proper Hadamard behavior.

\subsection{One loop two--point function}

In this subsection we basically repeat and generalize the calculations of \cite{PolyakovKrotov} and \cite{Leblond}. We would like to find the leading one loop contribution to $d^K(p\eta_1,\, p\eta_2)$ as $p\, \eta_{1,2}\to 0$ and $\eta_1/\eta_2 = const$ when we start from the vacuum state at the past infinity. Using Schwinger--Keldysh diagrammatic technic one obtains the leading answer:

\bqa\label{1loop}
d^K_1(p\eta_1, p\eta_2) \approx \frac12 \, h(p\eta_1)\, h^*(p\eta_2) \times \nonumber \\ \times \frac{\lambda^2}{2\, \pi^2}\, \int_p^{1/\eta} \frac{dk}{k}\, \iint_\infty^0 dx_1 \, dx_2 \left(x_1 \, x_2\right)^{\frac12} \, h\left[\frac{p}{k}\, x_1\right]\, h^*\left[\frac{p}{k}\, x_2\right]\, h^2(x_1)\, \left[h^*(x_2)\right]^2 - \nonumber \\ - h(p\eta_1)\, h(p\eta_2)\times \nonumber \\ \times \frac{\lambda^2}{2\,\pi^2}\, \int_p^{1/\eta} \, \frac{dk}{k} \, \int_\infty^0 dx_1 \,\int_\infty^{x_1} dx_2 \left(x_1 \, x_2\right)^{\frac12} \, h^*\left[\frac{p}{k}\, x_1\right]\, h^*\left[\frac{p}{k}\, x_2\right]\, h^2(x_1)\, \left[h^*(x_2)\right]^2 + c.c.
\eqa
where under the $k$ integral it is assumed that $1/\eta \equiv 1/\sqrt{\eta_1\eta_2} \gg k \gg p$. This formula reduces to the one obtained in \cite{PolyakovKrotov} when $\eta_1=\eta_2$.

From the last expression we see that although we have started from the vacuum state, where
$\left\langle a_p \, a_{-p}\right\rangle = 0$ and $\left\langle a^+_p \, a_p\right\rangle = 0$, these quantities are generated at one loop level. This can be traced back to the pair creation in dS space. Indeed in the course of evolution towards future infinity $\eta\to 0$ the density of the created particles appears to be:

\bqa
n_p(\eta) \equiv \langle a^+_p \, a_p\rangle = \frac{\lambda^2}{4\, \pi^2}\, \int_p^{1/\eta} \frac{dk}{k}\, \iint_\infty^0 dx_1 \, dx_2 \left(x_1 \, x_2\right)^{\frac12} \, h\left[\frac{p}{k}\, x_1\right]\, h^*\left[\frac{p}{k}\, x_2\right]\, h^2(x_1)\, \left[h^*(x_2)\right]^2.
\eqa
At the same time the anomalous quantum average is given by

\bqa
\kappa_p(\eta) \equiv \langle a_p \, a_{-p}\rangle = - \frac{\lambda^2}{2\,\pi^2}\, \int_p^{1/\eta} \, \frac{dk}{k} \, \int_\infty^0 dx_1 \,\int_\infty^{x_1} dx_2 \left(x_1 \, x_2\right)^{\frac12} \, h^*\left[\frac{p}{k}\, x_1\right]\, h^*\left[\frac{p}{k}\, x_2\right]\, h^2(x_1)\, \left[h^*(x_2)\right]^2.
\eqa
As the side remark let us point out that if one by mistake were using the standard Feynman diagrammatic technic in the circumstances under consideration, he would never see the appearance of the terms proportional to $h(p\eta_1)\, h(p\eta_2)$ or $h^*(p\eta_1)\, h^*(p\eta_2)$ in loop contributions to the Wightman function\footnote{From the Wightman function one can construct Keldysh and/or Feynman propagator.}. It is not hard to see that only terms proportional to $h^*(p\eta_1)\, h(p\eta_2)$ are generated in the loops within the standard Feynman technic. Furthermore in the stationary flat space case the Schwinger--Keldysh technic gives the same answer as the Feynman one, which reveals itself through the cancelation of all terms contributing to $\kappa_p$.

Now let us calculate the leading one loop IR contributions for various choices of harmonics $h(p\eta)$. To start we choose BD harmonics (\ref{BD}). For this choice, the $x_1$ and $x_2$ integrals in \eq{1loop} rapidly converge and are saturated around $x_{1,2}\sim \mu$, because in this case $h(x)\sim e^{ix}$ as $x\to \infty$. Hence, because $p/k\ll 1$ we can Taylor expand the $h(px/k)$ functions around zero, using their leading IR behavior

\bqa
h(x) \approx A_+ \, x^{i\mu} + A_-\, x^{-i\mu}, \quad x\to 0, \nonumber \\
A_+ = \frac{\sqrt{\pi}\,e^{\frac{\pi\mu}{2}}}{2^{i\mu + \frac12}\, \Gamma\left(1+i\mu\right)\, \sinh\left(\pi\mu\right)}, \quad A_- = \frac{\sqrt{\pi}\,e^{-\frac{\pi\mu}{2}}}{2^{-i\mu+\frac12}\, \Gamma\left(1-i\mu\right)\, \sinh\left(-\pi\mu\right)}.
\eqa
Furthermore, expanding $h(p\eta_{1,2})$ for small $p\eta_{1,2}$ and then keeping only leading IR terms in the $x_1$ and $x_2$ integrals, one can find the sum of the tree level and one loop leading IR contributions

\bqa\label{1result}
d^K_{0+1}\left(p\eta_1, p\eta_2\right) \approx \frac{\coth\left(\pi\mu\right)}{2\mu}\, s^{i\mu} + A_+ \, A^*_- \, \left(p\,\eta\right)^{2\, i\, \mu} \times \nonumber \\ \times \left\{1 + \frac{\lambda^2}{2\pi^2\mu}\, \log\left(\frac{1}{p\eta}\right)\, \iint^0_\infty dx_1 \, dx_2 \, \left(x_1 \, x_2\right)^{\frac12} \, h^2(x_1)\, \left[h^*(x_2)\right]^2 \times \right. \nonumber \\ \left. \times \left[\theta(x_1 - x_2)\, \left(\frac{x_1}{x_2}\right)^{i\mu} - \theta(x_2 - x_1) \left(\frac{x_1}{x_2}\right)^{-i\mu}\right]\right\} + c.c.,
\eqa
where $\theta(x_1-x_2)$ is the Heviside $\theta$--function and $s = \eta_1/\eta_2$, $\eta = \sqrt{\eta_1\eta_2}$. In deriving this result we have used that \cite{PolyakovKrotov}

\bqa
\left|\int_\infty^0 dx \, x^{\frac12 + i\, \mu}\, h^2(x)\right|^2 = e^{-2\, \pi\, \mu}\,\left|\int_\infty^0 dx \, x^{\frac12 - i\, \mu}\, h^2(x)\right|^2.
\eqa
Our result \eq{1result} reduces to the one obtained in \cite{PolyakovKrotov} when $s=1$.
Thus, we see that contributions to $n_p \equiv \langle a_p^+ \, a_p \rangle$ and $\kappa_p \equiv \langle a_p \, a_{-p}\rangle$ are of the same order. As we explain in the next section this means that BD harmonics are not suitable to write the KE only for $n_p$.

Let us consider a different choice of harmonics --- e.g. so called Jost functions at future infinity

\bqa
h(x) = \sqrt{\frac{\pi}{\sinh(\pi\mu)}}\, J_{i\mu}(x).
\eqa
They behave as $h(x) \sim x^{i\mu}$, when $x\to 0$. If one substitutes these harmonics into (\ref{1loop}), he should take into account that these harmonics behave at past infinity as

\bqa
h(x) = \sqrt{\frac{\pi}{4\, \sinh(\pi\, \mu)\, x}}\, \left[e^{i\,x} + e^{-\pi\,\mu - i\,x}\right], \quad x\to \infty.
\eqa
Hence, the $x_1$ and $x_2$ integrals in (\ref{1loop}) have contributions  around infinity due to the interference terms between $e^{ix}$ and $e^{-ix}$. They do not converge fast enough: they are saturated in the vicinity of $px/k\sim \mu$ rather than at $x\sim \mu$. Hence, naively one can not Taylor expand $h(px/k)$ around zero inside the $x_{1,2}$ integrals. However, let us subtract from and then add to $h^2(x)$ and $[h^*(x)]^2$ under the $x_{1,2}$ integrals the values of the interference terms, $\frac{\pi \, e^{-\pi\mu}}{4\, \sinh(\pi\, \mu)\, x}$:

$$h^2(x) = h^2(x) - \frac{\pi \, e^{-\pi\mu}}{4\, \sinh(\pi\, \mu)\, x} + \frac{\pi \, e^{-\pi\mu}}{4\, \sinh(\pi\, \mu)\, x}.$$
Then the $x_{1,2}$ integrals of $h^2(x) - \frac{\pi \, e^{-\pi\mu}}{4\, \sinh(\pi\, \mu)\, x}$ and of its complex conjugate are saturated around $x\sim \mu$ and one can Taylor expand $h(px/k)$ around zero inside the corresponding expressions. At the same time the contributions from the additional integrals of $\frac{\pi \, e^{-\pi\mu}}{4\, \sinh(\pi\, \mu)\, x}$ are suppressed in the IR limit. Indeed, due to extra powers of $q$ the momentum integrals, $dq$, are not divergent in the limit $p\eta\to 0$.

Thus, the leading IR contribution to the two--point function in this case is as follows:

\bqa\label{0+1}
d^K_{0+1}\left(p\eta_1, p\eta_2\right) \approx \frac{1}{2\, \mu}\, \left[s^{i\mu} + s^{-i\mu}\right]\, \left\{1 + \frac{\lambda^2}{2\,\pi^2\,\mu}\, \log\left(\frac{1}{p\eta}\right)\, \left|\int_\infty^0 dx \, x^{\frac12 + i\, \mu}\, \left[h^2(x) - \frac{\pi \, e^{-\pi\mu}}{4\, \sinh(\pi\, \mu)\, x}\right]\right|^2\right\}.
\eqa
We see that for this choice of the harmonics the particle number density $n_p$ dominates over $\kappa_p$ in the extreme IR limit. As we explain in the next section this means that the harmonics under consideration are suitable for the derivation of the KE in dS space at future infinity.

\subsection{de Sitter vs. Sphere QFT}

Before going on with KE
let us make a few comments about the possibility to formulate the dS QFT via analytical continuation from the sphere \cite{Marolf:2010zp}. The IR limit of the tree--level propagator for the BD harmonics is:

\bqa
d^K_0\left(p\eta_1,p\eta_2\right) \approx \frac{\coth\left(\pi\mu\right)}{2\mu}\, s^{i\mu} + A_+ \, A^*_- \, \left(p\,\eta\right)^{2\, i\, \mu} + c.c..
\eqa
Mass renormalization $\mu + \Delta \mu$ would lead to the one loop contribution of the following form:

\bqa
d^K_1\left(p\eta_1,p\eta_2\right) = \frac{\coth\left(\pi\mu\right)}{2\mu}\, s^{i\mu}\, i\, \Delta\mu\, \log(s) + A_+ \, A^*_- \, \left(p\,\eta\right)^{2\, i\, \mu}\, 2\, i\, \Delta\mu \, \log(p\eta) + c.c..
\eqa
Note that one expects the correction $\Delta \mu$ to be complex \cite{Leblond}.
The last expression can not reduce to the actual result (\ref{1result}) for any choice of $\Delta\mu$. In the light of what we have been saying in the sections above it is conceptually misleading to interpret the one loop contribution to the Wightman or Keldysh propagator in dS space as the mass renormalization, because its proper interpretation is in terms of particle creation --- in terms of slow functions $n_p$ and $\kappa_p$ of the average conformal time $\eta=\sqrt{\eta_1\,\eta_2}$.

At the same time the one loop Feynman diagrammatic technic calculation on the sphere can lead only to the mass renormalization: the authors of \cite{Marolf:2010zp} expected that after the appropriate subtraction of the UV divergences the remaining finite part would coincide with the one obtained by the direct calculation in dS space. However, now one can see that this can not happen. So it is not correct to define QFT on dS space via analytical continuation from the sphere, even in the situations when the tree--level dS correlation functions are indeed obtained via such a continuation.

Analytical continuation from the sphere may describe the stationary situation in dS, if any, but not the approach to the stationarity. The stationary situation corresponds to $\kappa_p=0$ and $n_p$ being independent of $\eta$. In fact, in the stationary situation the two--point correlation function depends only on the time difference (on $\eta_1/\eta_2$ in terms of the conformal time) rather than on both of the times ($\eta_1$ and $\eta_2$) independently. Such a situation can be achieved in dS space only in the extreme IR limit. From the one loop calculation above one may see that in BD state stationary situation can not happen in principal. However, for the Jost functions at future infinity the stationary situation may happen if the result of the summation of all loops --- $n_p(\eta)$ --- becomes somehow independent of time in the future infinity.

\section{The physical meaning of $\kappa_p$}

The one loop calculation of the previous section shows that the two--point function behaves as

\bqa
d^K(p\eta_1, p\eta_2) = \frac{1}{2} \, \left[h(p\eta_1)\, h^*(p\eta_2) + h^*(p\eta_1) \, h(p\eta_2)\right]\, \left[1 + 2\, n_p\left(\sqrt{\eta_1\eta_2}\right)\right] + \nonumber \\ + h(p\eta_1)\, h(p\eta_2)\,\kappa_p\left(\sqrt{\eta_1\eta_2}\right) + h^*(p\eta_1)\, h^*(p\eta_2)\,\kappa^*_p\left(\sqrt{\eta_1\eta_2}\right),
\eqa
when $p\eta_{1,2}\to 0$. Here $n_p(\eta)$ and $\kappa_p(\eta)$ are slow functions of their argument in comparison with the harmonics $h(p\eta)$. Note that they are  functions of time only, i.e. homogeneous in space. That is a natural situation in such a homogeneous space as dS, especially for the small values of the physical momenta.

Such a situation allows to write a system of KE for both $n_p \equiv \frac{1}{V}\, \left\langle a^+_p \, a_p \right\rangle$ and $\kappa_p \equiv \frac{1}{V}\, \left\langle a_p \, a_{-p} \right\rangle$. Here $V = \delta(0)$ is the comoving volume of the spatial sections. From the normalization of the harmonics it follows that then $n_p$ is the density per comoving volume --- per $V$. And $n_p$ appears in the expression for the total number density as follows: $N = \int d^3 p \, a^{-3}(\eta)\, n_p(\eta)$. Here $p$ is the comoving momentum, while $p/a(\eta)$ is the physical one. The particle number density per physical volume is defined as $\bar{n}_p = n_p/a^3(\eta)$.

The interaction Hamiltonian has the form (up to zero mode treatment, which is concisely discussed in the appendix):

\bqa
H_{int} = \frac{\lambda}{3}\, \int d^3k_1 \, d^3k_2 \, d^3k_3 \, a^4(\eta) \times \nonumber \\ \times \left\{3\, \delta\left(-\vec{k}_1 + \vec{k}_2 + \vec{k}_3\right) \, \left[g^*_{k_1}\, g_{k_2}\, g_{k_3}(\eta)\, a^+_{k_1}\, a_{k_2}\, a_{k_3} + g_{k_1}\, g^*_{k_2}\, g^*_{k_3}(\eta)\, a_{k_1}\, a^+_{k_2}\, a^+_{k_3}\right] \right. \nonumber \\ + \left. \delta\left(\vec{k}_1 + \vec{k}_2 + \vec{k}_3\right) \, \left[g_{k_1}\, g_{k_2}\, g_{k_3}(\eta)\, a_{k_1}\, a_{k_2}\, a_{k_3} + g^*_{k_1}\, g^*_{k_2}\, g^*_{k_3}(\eta)\, a^+_{k_1}\, a^+_{k_2}\, a^+_{k_3}\right]\right\},
\eqa
Using this Hamiltonian one can find the evolution in time of $\langle a^+a\rangle$ and $\langle aa \rangle$ and proceed along the same lines as in the first part of the Appendix. But now we have to take into account, when perform Wick contractions, that not only $\left\langle a^+_p \, a_{p'} \right\rangle = n_p \, \delta(p-p')$, but as well $\left\langle a_p \, a_{p'} \right\rangle = \kappa_p \, \delta(p+p')$ and $\left\langle a^+_p \, a^+_{p'} \right\rangle = \kappa_p^* \, \delta(p+p')$.

With the use of the following matrixes:

\bqa
N_p(\eta_1,\, \eta_2) = \left(
  \begin{array}{cc}
    n_p(\eta_1) \, g^*_p(\eta_2) & \kappa_p(\eta_1) \, g_p(\eta_2) \\
    \kappa_p(\eta_1) \, g_p(\eta_2) & n_p(\eta_1) \, g^*_p(\eta_2) \\
  \end{array}
\right), \quad P = \left(
  \begin{array}{cc}
    0 & 1 \\
    1 & 0 \\
  \end{array}
\right)
\eqa
the resulting system of KE can be written in a compact form. The real equation has the form:

\bqa\label{callintPP0}
\frac{d n_p(\eta)}{d\eta} = \left[N\to P\,N\right] + 2\,\lambda^2 \, \int \frac{d^3k_1\, d^3k_2}{(2\pi)^6} \, \int_{\eta_0}^{\eta} \frac{d\eta'}{\left(\eta\, \eta'\right)^4} \, \, {\rm Re}\, \, Tr \nonumber \\ \left\{\delta\left(\vec{p} - \vec{k}_1 - \vec{k_2}\right)\, C_{pk_1k_2}(\eta)\, \left[\left(1+N^*_p\right)\, N_{k_1}\, N_{k_2}^{\phantom{\frac12}} - \,\, N^*_p \, \left(1+N_{k_1}\right)\, \left(1+N_{k_2}\right)\right](\eta',\eta') +  \right. \nonumber \\ + 2\,\delta\left(\vec{k}_1 - \vec{k}_2 - \vec{p}\right) \, C_{k_1k_2p}(\eta)\, \left[ N^*_{k_1}\, \left(1+N_{k_2}\right)\,\left(1+N_p\right)^{\phantom{\frac12}} - \,\,\left(1+N^*_{k_1}\right)\, N_{k_2}\, N_p \right](\eta',\eta') + \nonumber \\ \left.
+ \delta\left(\vec{p} + \vec{k}_1 + \vec{k_2}\right)\, D_{pk_1k_2}(\eta)\,\left[ \left(1+N_{p}\right)\, \left(1+N_{k_1}\right) \,\left(1+N_{k_2}\right)^{\phantom{\frac12}} - \,\, N_{p}\, N_{k_1}\, N_{k_2} \right](\eta',\eta')^{\phantom{\frac12}}\right\},
\eqa
while the complex one is as follows:

\bqa\label{callintPP00}
\frac{d \kappa_p(\eta)}{d\eta} = \left[N\to P\,N\right] + 2\,\lambda^2 \, \int \frac{d^3k_1\, d^3k_2}{(2\pi)^6} \, \int_{\eta_0}^{\eta} \frac{d\eta'}{\left(\eta\, \eta'\right)^4} \, \, \left(\vec{p}\to - \vec{p}\right)\, \, Tr \nonumber \\ \left\{\delta\left(\vec{p} - \vec{k}_1 - \vec{k_2}\right)\, C_{pk_1k_2}(\eta)\,\left[ \left(1+N_{p}\right)\, \left(1+N_{k_1}\right) \,\left(1+N_{k_2}\right)^{\phantom{\frac12}} - \,\, N_{p}\, N_{k_1}\, N_{k_2} \right](\eta',\eta') +  \right. \nonumber \\ + 2\,\delta\left(\vec{k}_1 - \vec{k}_2 - \vec{p}\right) \, C^*_{k_2k_1p}(\eta)\, \left[ N^*_{k_1}\, \left(1+N_{k_2}\right)\,\left(1+N_p\right)^{\phantom{\frac12}} - \,\,\left(1+N^*_{k_1}\right)\, N_{k_2}\, N_p \right](\eta',\eta') + \nonumber \\ \left.
+ \delta\left(\vec{p} + \vec{k}_1 + \vec{k_2}\right)\, D^*_{pk_1k_2}(\eta)\, \left[\left(1+N_p\right)\, N^*_{k_1}\, N_{k_2}^{*\phantom{\frac12}} - \,\, N_p \, \left(1+N^*_{k_1}\right)\, \left(1+N^*_{k_2}\right)\right](\eta',\eta')^{\phantom{\frac12}}\right\},
\eqa
$\eta_0$ is the moment of time when we switch on the interactions. The notation $\left[N\to P\,N\right]$
means that we have to add to the explicitly written expression the same one where every $N$ is substituted by the product $P\,N$; Re $Tr$ means that one has to take the real part and the trace of the expression following after these signs; at the same time $\left(\vec{p}\to - \vec{p}\right)\,Tr$ means that one has to take the trace and add to the expression following after these signs the same one with the exchange $\vec{p}\to - \vec{p}$. Finally $C_{k_1k_2k_3} = g^*_{k_1}\,g_{k_2}\, g_{k_3}$ and $D_{k_1k_2k_3} = g_{k_1}\,g_{k_2}\, g_{k_3}$.

To convert (\ref{callintPP0}), \eq{callintPP00} into a tractable form one has to assume that $n_p(\eta')$ and $\kappa_p(\eta')$ are slow functions in comparison with the harmonics. The situation is similar to the one discussed in the Appendix. Then the first argument of the matrix $N_k(\eta',\eta')$ can be taken to be $\eta$ instead of $\eta'$. Furthermore, explicit check shows that one can extend the limits of the $\eta'$ integration inside the collision integral (CI) to $\eta\to 0$, $\eta_0\to \infty$. This does not make the $\eta'$ and $k$ integrals divergent.

Expanding the terms under the CI in \eq{callintPP0} one will encounter the standard expressions which appear in the CI written in the Appendix:

\bqa\label{term}
\left[(1+n_p)\, n_{k_1}\, n_{k_2} - n_p \, (1+n_{k_1})\, (1+n_{k_2})\right], \nonumber \\
\left[n_{k_1}\, (1+n_{k_2})\,(1+n_p) - (1+n_{k_1})\, n_{k_2}\, n_p \right], \nonumber \\
\left[(1+n_p)\, (1 + n_{k_1})\, (1 + n_{k_2}) - n_p \, n_{k_1}\, n_{k_2}\right]
\eqa
These three contributions have the following physical meaning. The first term describes the competition between two processes. One that the wave with the momentum $p$ decays into two waves $\vec{k}_1 + \vec{k}_2 = \vec{p}$. This process, corresponding to the term $n_p \, (1+n_{k_1})\, (1+n_{k_2})$, appears with the minus sign in the CI because it describes the loss of the wave with the momentum $p$. The inverse gain process, corresponding to the term $(1+n_p)\, n_{k_1}\, n_{k_2}$ with the plus sign, is that when two waves merge to create the wave with the momentum $p$.

The second term in \eq{term} describes as well two competing processes. The first process is that the wave with the momentum $p$ can merge together with another wave (with the momentum $k_2$) to create a third one (with the momentum $k_1$). This is the loss process. The inverse gain process happens when a wave (with the momentum $k_1$) decays in to two, one of which is with the momentum $p$. The coefficient 2 in front of this term is just the combinatoric factor because we can exchange $\vec{k}_1$ and $\vec{k}_2$.

The third term as well describes two processes.
The gain process is when three waves, one of which is with the momentum $p$,
are created out of vacuum. The loss process is when three waves are annihilated into the vacuum.
All these processes are not allowed by energy--momentum conservation for {\it massive}
particles in flat space. However, in dS space all these processes are allowed \cite{Myrhvold}, \cite{Boyanovsky:2004ph}, \cite{Bros:2009bz}, \cite{Volovik:2008ww}, \cite{Akhmedov:2009be}, \cite{AkRoSa}, \cite{Polyakov:2009nq}, because there is no energy conservation.

Due to the presence of $\kappa_p$ in \eq{callintPP0} we have extra terms in CI. All of them can be obtained from those listed in \eq{term} via the simultaneous substitutions of $(1+n_{k,p})$'s and $n_{k,p}$'s by $\kappa_{k,p}$'s or $\kappa^*_{k,p}$'s. E.g. we encounter terms of the form:

\bqa
\left[(1+n_p)\, \kappa_{k_1}\, n_{k_2} - n_p \, \kappa_{k_1}\, (1+n_{k_2})\right]
\eqa
which as well correspond to the two competing processes --- one that the wave with the momentum $p$ lost (gained) in such a process that instead of the creation (annihilation) of the two particles $k_1$ and $k_2$ we see single $k_2$ and the missing momentum $k_1$ is gone into (taken from) the background quantum state of the theory. Which should not be confused with the background geometry.

Similarly we encounter terms of the form

\bqa
\left[(1+n_p)\, \kappa_{k_1}\, \kappa_{k_2} - n_p \, \kappa_{k_1}\, \kappa_{k_2}\right]
\eqa
which describe the processes in which both $k_1$ and $k_2$ are coming from (going to) the background state.

Thus, we find it natural to interpret the anomalous quantum average $\kappa_p = \langle a_p \, a_{-p}\rangle$ as the measure of the strength of the backreaction on the background quantum state of the theory of the various processes described by the standard terms \eq{term} in CI. Then it should be expected that in the vicinity of the equilibrium the backreaction is small. Which means that $\kappa_p$ is suppressed in comparison with $n_p$ and can be neglected. In such a situation the system of KE reduces to one equation for $n_p$ only.

Rephrasing, if we start from an initial state, which is substantially different from the eventual
stationary one, the generated $\kappa_p$ is comparable to $n_p$.
But as the state of the theory approaches the equilibrium, $\kappa_p$ becomes suppressed, which signals the small backreaction.

The picture we have in mind is as follows. Note first of all that the in--state (say BD one) looks as the coherent state from the point of view of our preferable out--state (specified by the above out Jost harmonics). The situation is similar to the one for the QED in the constant electric field background \cite{GribMamaevMostepanenko}. Now consider a flat space ordinary QFT, which has the unique Poincare invariant vacuum state. Consider its evolution in time if the in--state is some excited coherent state. Intuitively it is natural to think that the final state of the theory will be build on the basis of the appropriate vacuum state under consideration. And one can explicitly see that in the in--state the anomalous quantum average $\langle aa \rangle$ will be of the same order as $\langle a^+ a\rangle$, while in the out--state the anomalous quantum average will be suppressed. Here $a^+$ and $a$ are creation and annihilation operators corresponding to the correct Poincare invariant vacuum state.

\section{Kinetic equation in Poincare patch}

Taking into account the one loop calculation above we conclude that if one starts from the BD state at past infinity then he has to solve KE equation for $n_p$ and $\kappa_p$ together.
At the same time, we may assume that in the expanding PP of dS space there is a final state, which is close to stationarity and is build on the basis of the Fock space corresponding to the out Jost harmonics $h(x)\sim J_{i\mu}(x)$. For the latter harmonics one can write the KE equation containing only $n_p$.

Then, as $\eta\to 0$, we can put $\kappa_p = 0$ in \eq{callintPP0} to arrive at

\bqa\label{callintPP}
\frac{d n_p(\eta)}{d\log (\eta)} = \frac{\lambda^2}{\pi^2} \, \int_0^{\infty} dk\, \eta \,(k\eta)^{\frac12} \, \int_{\infty}^{0} dy' \, \left(y'\right)^{\frac12}\times \nonumber \\ \times \left\{ {\rm Re} \left(C\left[\frac{p}{k}\,k\eta,\, k\eta,\, \frac{|p-k|}{k}\,k\eta\right] \, C^*\left[\frac{p}{k}\,y',\, y',\,\frac{|p-k|}{k}\,y'\right]\right) \times \right.\nonumber \\ \times \left[(1+n_p)\, n_{k}\, n_{|p-k|}^{\phantom{\frac12}} - \,\, n_p \, (1+n_{k})\, (1+n_{|p-k|})\right](\eta) + \nonumber \\ + 2\,{\rm Re} \left(C\left[k\eta,\, \frac{|k-p|}{k}\, k\eta,\, \frac{p}{k}\,k\eta\right] \, C^*\left[y',\, \frac{|k-p|}{k}\,y',\,\frac{p}{k}\, y'\right] \right)\times \nonumber \\ \times \left[ n_{k}\, (1+n_{|k-p|})\,(1+n_p)^{\phantom{\frac12}} - \,\,(1+n_{k})\, n_{|k-p|}\, n_p \right](\eta) + \nonumber \\
+ {\rm Re} \left(D\left[k\eta,\, \frac{|p+k|}{k}\,k\eta,\, \frac{p}{k}\,k\eta\right]\, D^*\left[y',\,\frac{|p+k|}{k}\, y',\, \frac{p}{k}\, y'\right]\right) \times \nonumber \\ \times \left. \left[ (1+n_{k})\, (1+n_{|p+k|}) \,(1+n_p)^{\phantom{\frac12}} - \,\, n_{k}\, n_{|p+k|}\, n_p \right](\eta)^{\phantom{2}}\right\},
\eqa
where $C[x,y,z] = h^*(x)\, h(y)\, h(z)$, $D[x,y,z] = h(x)\, h(y)\, h(z)$ and $h(x)$ is the specified above set of solutions of the Bessel equation. In deriving this equation we assumed that all quantities under the integral over $d^3\vec{k}$ depend only on $\left|\vec{k}\right|$ and changed the variables $y' = k\eta'$.

This is not yet the equation we are looking for. But before going on let us make a few comments. First, it is not hard to see that for short periods of time $t = - \log \eta\ll 1$, i.e. well within the cosmological horizon, and for $m,k\gg 1$, the $\eta'$ integrals simplify and the above equation reduces to the one in flat space presented in the Appendix. However, for cosmological times one has to use solutions of the Bessel equation instead of the flat space plain waves. Then one does not obtain the $\delta$--functions ensuring energy conservation inside the CI. That is of cause should be the case in such a time dependent gravitational background as dS space.

Second, the prefactors of the $n_p$ dependent terms inside the CI can be considered as the definitions of the rates of the six processes described by the CI. This can be a way to define the rates in the circumstances when there is no well defined notion of the S--matrix \cite{Akhmedov:2008pu}.

For the small physical momenta ($p\eta \to 0$) one should look for a solution of this equation in the form of the function of the physical momentum $p\,\eta$ --- $n_p(\eta) = n(p\eta)$. In fact, PP brakes some part of dS isometry, but is invariant at least under the simultaneous
rescaling of $\eta$ and $\vec{x}$. Hence, even if we start with the non--invariant under this symmetry state one can expect that due to the expansion, which smoothes everything out, unless backreaction becomes strong, all physical quantities are going to be invariant under the symmetry in question in the future infinity.
In particular, that means that the distribution $n_p(\eta)$ should depend on the invariant quantity $p\,\eta$.

Indeed, for the small physical momenta the equation (\ref{callintPP}) reduces to:

\bqa\label{callintPP2}
\frac{d n(x)}{d\log (x)} = \frac{\lambda^2}{4\,\pi^2\, \mu} \, \int_0^{\infty} dy \,(y)^{\frac12} \, \int_{\infty}^{0} dy' \, \left(y'\right)^{\frac12} \times \nonumber \\ \times \left\{ {\rm Re} \left\{y^{-i\mu} \, \left[h^2\left(y\right) - \frac{\pi \, e^{-\pi\mu}}{4\, \sinh(\pi\, \mu)\, |y|}\right] \,\left(y'\right)^{i\mu}\, \left[\left(h^*\left(y'\right)\right)^2- \frac{\pi \, e^{-\pi\mu}}{4\, \sinh(\pi\, \mu)\, \left|y'\right|}\right]\right\} \times \right. \nonumber \\ \times \left[(1+n(x))\, n(y)^{2\phantom{\frac12}} - \,\, n(x) \, (1+n(y))^2\right] + \nonumber \\
+ 2\,{\rm Re} \left\{y^{i\mu} \, \left[\left|h\left(y\right)\right|^2- \frac{\pi \, e^{-\pi\mu}}{4\, \sinh(\pi\, \mu)\, |y|}\right] \, \left(y'\right)^{- i\mu} \, \left[\left|h\left(y'\right)\right|^2- \frac{\pi \, e^{-\pi\mu}}{4\, \sinh(\pi\, \mu)\, \left|y'\right|} \right]\right\} \times \nonumber \\ \times \left[ n(y)\, (1+n(y))\,(1+n(x))^{\phantom{\frac12}} - \,\,(1+n(y))\, n(y)\, n(x) \right] + \nonumber \\
+ {\rm Re} \left\{y^{i\mu} \, \left[h^2\left(y\right) - \frac{\pi \, e^{-\pi\mu}}{4\, \sinh(\pi\, \mu)\, |y|}\right] \, \left(y'\right)^{- i\mu} \, \left[\left(h^*\left(y'\right)\right)^2- \frac{\pi \, e^{-\pi\mu}}{4\, \sinh(\pi\, \mu)\, \left|y'\right|}\right]\right\} \times \nonumber \\ \times \left. \left[ (1+n(y))^2 \,(1+n(x))^{\phantom{\frac12}} - \,\, n(y)^2\, n(x) \right]^{\phantom{2}}\right\},
\eqa
where $x=p\eta$, $y=k\eta$ and $y' = k\eta'$.
In the last expression we have neglected the contribution from the region $k<p$ to the integral over $k$ (i.e. over $y$), because the main contribution to the integrals comes from $k\, \eta \sim \mu$, while $p\,\eta \ll \mu$.

\section{Solution in the expanding Poincare patch}

As we explain in the Appendix in Minkowski space--time the Plankian distribution $n_p = 1/(e^{\omega/T} - 1)$ annihilates the first two contributions inside the CI (\ref{callintPP}). The last term in the CI is just forbidden by the energy conservation in Minkowski space.  However, neither the first two terms nor the last term in \eq{callintPP} are annihilated by the Plankian distribution, because there is no any restriction on the energy in dS space. Note that even for the case $n_p=0$ the CI does not vanish due to the particle creation out of vacuum.

Now we assume that in the future infinity of the expanding PP, or for the low physical momenta,
$n(p\eta)$ is very small. At the same time the particle number density for the high physical momenta should be even smaller, $n(x) \gg n(y)$ for $y\gg x$. Hence, in this limit one can approximate

\bqa
(1+n(x))\, n(y)^2 - n(x) \, (1+n(y))^2 \approx - n(x),\nonumber \\
n(y)\, (1+n(y))\,(1+n(x)) - (1+n(y))\, n(y)\, n(x) \approx n(y), \nonumber \\
(1+n(y))^2 \,(1+n(x)) - n(y)^2\, n(x) \approx 1.
\eqa
Furthermore, the main contribution to the second term \eq{callintPP2} comes from the region $y\sim \mu$, while $x\ll \mu$. Hence, we can neglect the second term on the RHS of (\ref{callintPP2}) because it is proportional to $n(y\sim \mu)\ll n(x)$.
Then the KE reduces to:

\bqa\label{solution1}
\frac{d n(x)}{d\log (x)} = \Gamma \, n(x) - \Gamma', \nonumber \\
\Gamma = \frac{\lambda^2}{2\pi^2\, \mu} \, \left|\int_0^{\infty} dy \,y^{\frac12 - i\, \mu} \, \left[h^2\left(y\right) - \frac{\pi \, e^{-\pi\mu}}{4\, \sinh(\pi\, \mu)\, |y|}\right] \right|^2, \nonumber \\ \Gamma' = \frac{\lambda^2}{2\,\pi^2\, \mu} \, \left|\int_0^{\infty} dy \,y^{\frac12 + i\,\mu} \, \left[h^2\left(y\right) - \frac{\pi \, e^{-\pi\mu}}{4\, \sinh(\pi\, \mu)\, |y|}\right]\right|^2.
\eqa
Which is a kind of renormalization group equation where the CI plays the role of the
$\beta$--function \cite{Polyakov:2009nq}. In fact, we obtain an equation which shows how the distribution of particles $n(p\eta)$
changes with the change of the scale $p\eta$. Moreover, as follows from the discussion in the Appendix the solution of the KE sums up bubble diagrams, which is exactly what the renormalization group equations do.
Similar situation is discussed in \cite{Burgess:2010dd}.

The solution of the obtained differential equation is

\bqa\label{solu}
n(p\eta) = \frac{\Gamma'}{\Gamma}\left[C \, \left( p\,\eta\right)^{\Gamma} + 1\right],
\eqa
where $C$ is the integration constant, which may depend on the initial conditions. The obtained solution is valid because $\frac{\Gamma'}{\Gamma} \approx e^{-2\,\pi\mu}\ll 1$ for $\mu\gg 1$. Furthermore, at the leading order in $\lambda^2$ we reproduce the one loop result of the previous section if $C=-1$.
One should recall here that $n(p\eta)$ is the particle density per comoving volume.

The obtained distribution has the stationary value $e^{-2\pi\mu}$, which approximately annihilates the CI in the IR limit. It is reached when the production of particles is equilibrated by their decay.
In fact, from what we have mentioned above
it is clear that $\Gamma$ defines the decay rate of the scalar particle into two, while $\Gamma'$ defines the particle production rate. Towards the future infinity $\log x$ is decreasing, hence, indeed $\Gamma'$ corresponds to the gain, while $\Gamma$ to the loss in (\ref{solution1}). Furthermore, obviously the loss in question should be proportional to the particle density, while the gain should be just a constant for low distributions.

At the stationary point in question the theory in dS space can be described by the analytical continuation from the sphere. But one can not describe the approach to the equilibrium \eq{solu} via such an analytical continuation.

Finally, we have obtained the solution \eq{solu} assuming that $n(p\eta)$ decays in the future due to the expansion of the space and despite the constant rate of particle production. Such an assumption is natural if we start with low particle density at past infinity. However, there is a question if this situation would change or not once we will start with some different state at past infinity?
Of cause if we will start with a very high density of particles, in comparison with the vacuum energy, then the situation will be much different, but what will happen say at the intermediate densities?
To address this issue one may look for other solutions of the KE \eq{callintPP}. But we find it
more appropriate to study the problem in global dS space, because
global dS sets up initial conditions for the PP --- for the inflation. This will be done elsewhere.

\section{Solution in contracting Poincare patch}

In the {\it contracting} PP we do not expect stationary situation in the future infinity \cite{PolyakovKrotov}. Hence, we do not expect KE to be applicable in this situation. But for the moment let us assume that the KE is applicable and see what kind of solution it leads to. We will explicitly see that for the obtained solution the approximation, which we have used to derive the KE, brakes down for late enough times.

It is not hard to obtain the KE for the low physical momenta in the contracting patch via direct derivation or via the time--reversal ($t\leftrightarrow-t$ or $\eta \leftrightarrow 1/\eta$) from the equation in the expanding patch. The only change wrt the equation (\ref{callintPP}) is the relative sign between the LHS and RHS. For low physical momenta in the contracting PP we use $h(x)\sim Y_{i\mu}(x)$, where $Y$ is the Bessel function of the second type, but now it plays the role of the Jost functions at past infinity.

We are interested in the behavior of the solution at the future infinity $\eta \to \infty$ and expect that for low momenta the distribution grows with time, due to the contraction of the space and constant particle production. As we will see in a moment the products of $C[x,y,z]$ and $D[x,y,z]$ on their complex conjugates in (\ref{callintPP}) do not depend on $p$ and $k$ for low enough $x,y$ and $z$. That is related to the usual flatness of the spectrum in dS background.

Hence, it is natural to assume that $n_p(\eta)$ for $p\eta \ll \mu$ does not depend on $p$ and can be taken out from the integral over $k$ on the RHS of (\ref{callintPP}). Moreover, due to the expected explosive growth of $n(\eta)$ for the low momenta we can make the following approximations

\bqa
(1+n_p)\, n_{k}\, n_{p-k} - n_p \, (1+n_{k})\, (1+n_{p-k}) \approx - n^2(\eta), \nonumber \\
n_{k}\, (1+n_{k-p})\,(1+n_p) - (1+n_{k})\, n_{k-p}\, n_p \approx n^2(\eta), \nonumber \\
(1+n_{k})\, (1+n_{p+k}) \,(1+n_p) - n_{k}\, n_{p+k}\, n_p \approx n^2(\eta)
\eqa
if $k$ and $p$ are small enough.

The reason why the range of momenta for which our approximation is valid ($p\eta\ll \mu$) narrows down as the time goes by ($\eta\to \infty$) is as follows. Due to the contraction of the space in question the long wave length fluctuations cross into the horizon with the growth of time. But there are the horizon size modes which show the explosive behavior of their distribution.

Having made these observations, let us split the integrals over $y$ and $y'$ in the CI
into two parts: one due to the region where $y,y'\ll \mu$ and the other due to the region where $y,y'\gg \mu$. Then the KE reduces to:

\bqa\label{callintPP3}
\frac{d n(\eta)}{d\log (\eta)} = \frac{\lambda^2}{4\,\pi^2\, \mu^3} \, \int_0^{\mu} dy \,(y)^{\frac12} \, \left\{ -{\rm Re} \left[y^{-i\, \mu} \int^{\mu}_{0} dy' \,(y')^{\frac12 + i\, \mu}\right] \, n^2(\eta)\right. + \nonumber \\
+ 2\,{\rm Re} \left[y^{-i\,\mu}\int^{\mu}_{0} dy' \, (y')^{\frac12 + i\, \mu}\right]\, n^2(\eta) + \nonumber \\
+ \left. 3\,{\rm Re} \left[y^{-3\,i\, \mu}\int^{\mu}_{0} dy'\, (y')^{\frac12 + 3\, i\, \mu} \right] \, n^2(\eta)\right\} + \dots
\eqa
Dots on the RHS of this equation stand for the terms which are suppressed in comparison
with the terms explicitly written in this equation, because $n(y)$ for $y \gg \mu$ should be small as compared to $n(\eta)$. Thus the integro--differential KE again reduces to the differential equation of the renormalization group type:

\bqa
\frac{dn(\eta)}{d\log(\eta)} = \bar{\Gamma} \, n^2(\eta), \quad {\rm where} \quad
\bar{\Gamma} \approx \frac{\lambda^2 \, \mu^2}{2\, \pi^2\, m^2\, \left(m^2 - \frac32\right)} > 0.
\eqa
The solution of this equation is:

\bqa
n(\eta) \sim \frac{1}{A - \bar{\Gamma}\, \log\eta} \sim \frac{1}{\bar{\Gamma}\,\log\frac{\eta_0}{\eta}},
\eqa
and is indeed independent of $p$. The solution under consideration is valid for $\eta < \eta_0 = e^{const/\lambda^2}\gg 1$, $A$ is an integration constant, which may depend on the previous history of the evolution of the state and, in particular, on the initial state.

As we see this solution has a pole at some finite $\eta_0$. The situation is similar to the renormalization group Landau--Pomeranchuk pole if we consider the CI as the $\beta$--function. In the vicinity of the point $\eta_0$ the approximation in which we have derived the KE brakes down, but in any case we already have to take into account backreaction on the background.

\section{Conclusions}

Several comments are in order at this point.
First, in this note we have considered 4D $\phi^3$ QFT, but most of our conclusions and formulas can be easily extended to the arbitrary dimensions and for $\phi^4$ theory.
As well one can write the CI at higher orders in $\lambda$. In fact, from the discussion in the main body of the text the structure and physical origin of all terms in the CI should be clear: one should just include into CI all possible processes, which are allowed by the momentum conservation, and substitute the plain waves $e^{i\,\omega\, t}$ (inside the expressions leading to the $\delta$--functions establishing energy conservation) by the out Jost harmonics.

Second, the term in the CI (\ref{callintPP}), which is responsible for the creation of particles out of the vacuum, does not vanish as well for the odd dimensional dS space--times.
Thus, there is the particle production also in the odd dimensional dS space--times. This observation contradicts some earlier climes in the literature.

In conclusion, we have shown that the proper interpretation of the Schwinger--Keldish one loop renormalization of the two point function is in terms of generation of the slow functions $n_p = \langle a^+_p\, a_p \rangle$ and $\kappa_p = \langle a_p\, a_{-p}\rangle$. Which signals the particle creation in dS space. We have explicitly shown that the one loop contribution to the two--point function does not reduce to the mass renormalization. Hence, loop corrections for the quantum fields in dS space can not be obtained via analytical continuation from the sphere.

We have shown that for the BD harmonics the IR value of $\kappa_p$ is of the same order as $n_p$. Hence, these harmonics are not suitable for writing KE for $n_p$ only: For BD state one has to solve the system of KE for both $n_p$ and $\kappa_p$. We derive such a system of KE as well. However, for the Jost harmonics $g_p(\eta) \sim \eta^{3/2}\, J_{i\mu}(p\,\eta)$ the IR value of $\kappa_p$ is suppressed in comparison with the one for $n_p$. We suspect that this observation means that from whatever homogeneous state at past infinity in PP we start eventually the state of the theory flows to the one build with the use of the future Jost harmonics. In fact, as we explain in the main body of the text, it is natural to interpret the anomalous quantum average $\kappa_p = \langle a_p \, a_{-p}\rangle$ as the measure of the strength of the backreaction on the background quantum state of various processes described by the standard terms in CI.
Thus, if we start from an initial state, which is substantially different from the eventual
stationary one, the generated $\kappa_p$ is comparable to $n_p$. But as the state of the theory approaches the equilibrium, $\kappa_p$ becomes suppressed, which signals the small backreaction.

We have derived the KE containing only $n_p$
both in expanding and contracting PP of dS space and found solutions of these equation.
These solutions can be understood on general physical grounds.
In the expanding PP it looks like

\bqa
n(p\eta) = \frac{\Gamma'}{\Gamma}\left[C \, \left( p\,\eta\right)^{\Gamma} + 1\right],
\eqa
where $C$ is the integration constant, which may depend on the initial conditions. The obtained solution is valid because $\frac{\Gamma'}{\Gamma} \approx e^{-2\,\pi\mu}\ll 1$ for $\mu\gg 1$.
It has a stationary Gibbons--Hawking value $e^{-2\pi\mu}$, which approximately in IR limit annihilates the CI and is reached when the production of particles is equilibrated by their decay.
In fact, from what we have stated in the main body of the text it is clear that $\Gamma$ defines the decay rate of the scalar particle into two, while $\Gamma'$ defines the particle production rate.

Note that $n_p(\eta)$ is the density per comoving volume. The density per physical volume decays to zero in the future infinity because of the expansion of the physical volume: $\bar{n}_p \approx \eta^3 \,\frac{\Gamma'}{\Gamma} \, \left[C\,\left(p\, \eta\right)^{\Gamma} + 1\right]$ (see the discussion above).

In the contracting PP the solution for low momenta $p$ is

\bqa
n(\eta) \sim \frac{1}{A - \bar{\Gamma}\, \log\eta} \sim \frac{1}{\bar{\Gamma}\,\log\frac{\eta_0}{\eta}},
\eqa
and is independent of $p$. The solution under consideration is valid for $\eta < \eta_0 = e^{const/\lambda^2}\gg 1$, $A$ is an integration constant, which may depend on the previous history of the evolution of the state and, in particular, on the initial state.
As expected the distribution grows with time, due to the contraction of the space and constant particle production, and moreover has a pole at some finite $\eta_0$.

The use of the expression ``particle density'' in this context demands some clarifications. As we mentioned in the main body of the text one can not diagonalize the free hamiltonian $H_0$ in dS space once
and forever. One can find harmonics which make $H_0$ diagonal at past infinity, but there are no harmonics which diagonalize $H_0$ in the future, because $B_k$ is time dependent as $\eta\to 0$. To deal with the appropriate notion of particles it is tempting to find such harmonics which diagonalize the free Hamiltonian instantaneously at a fixed moment of time $\eta$.
This can be achieved via the time dependent Bogolyubov transform \cite{GribMamaevMostepanenko}: $
b_k(\eta) = \alpha_k(\eta) \, a_k + \beta_k(\eta) \, a_{-k}^+, \quad
b^+_k(\eta) = \alpha^*_k(\eta) \, a^+_k + \beta^*_k(\eta)\, a_{-k}$. The harmonics are

\bqa\label{Bogol}
\bar{g}_k(\eta) = \alpha_k^* \, g_k - \beta^*_k\, g^*_k = \frac{i}{a(\eta)\,\left[k^2 + a^2(\eta)\, m^2\right]^{\frac14}}\, \frac{\frac{dg_k}{d\eta} - i\, \sqrt{k^2 + a^2(\eta)\,m^2}\, g_k}{\left|\frac{dg_k}{d\eta} - i\, \sqrt{k^2 + a^2(\eta)\,m^2}\, g_k\right|}.
\eqa
Hence, it is tempting to derive the KE for $\langle b^+_p\, b_p\rangle$ --- appropriate particle density at the given moment of time.

However, because we have made a time dependent canonical (Bogolyubov) transformation to arrive at these harmonics we have to take into account the explicit time dependence of $b$'s. Then the problem is that the evolution equation closes wrt $\langle b^+_p\, b_p\rangle$ only if $\dot{\alpha}_p\, \beta_p - \alpha_p \, \dot{\beta}_p$ is negligible in comparison with the CI. However, this is not the case in dS space. In the latter case the corresponding equation does not even close wrt both $\langle b^+b\rangle$ and $\langle bb\rangle$. Hence, in dS space it is more appropriate to use the above Jost harmonics.
Furthermore, if one knows $\langle a^+ a\rangle$, $\langle a a \rangle$ and $\langle a^+ a^+ \rangle$ as the solutions of the KE, then he can find $\langle b^+ b\rangle$, $\langle b b \rangle$ and $\langle b^+ b^+ \rangle$ via Bogolyubov transformation (\ref{Bogol}). But once $\langle a\, a\rangle = \langle a^+ \, a^+\rangle = 0$ for some choice of the harmonics in the IR limit, then these harmonics can be considered as the proper quasiparticles and there is no need to look for $b$ and $b^+$.

\section{Acknowledgements}

I would like to acknowledge (in random order) discussions with A.Dymarsky, O.Kancheli, N.Afshordi, P.Buividovich, M.Voloshin, A.Kamenev, G.Baskaran, L.Leblond, D.Krotov, A.Semenov, P.Arseev, A.Morozov, A.Mironov, V.Losyakov, S.Apenko, E.Mottola, U.Moschella, I.Bars, L.Boyle, M.Johnson, A.Albrecht and I.Burmistrov. Especially I would like to thank A.Roura for the very valuable comments and objections. As well I would like to thank P.Burda and A.Sadofyev for the collaboration at the early stage of the work on this project. Finally, I would like to thank Perimeter Institute and Albert Einstein Institute for Gravitational Physics for the hospitality during the most productive part of the period when this work was done. The work was done under the partial financial support of grants for the Leading Scientific Schools NSh-6260.2010.2 and RFBR 08-02-00661-a.

\section{Appendix}

To make the paper self--contained in this Appendix we present the derivation of the KE
for the flat space massless real scalar field theory with the cubic self--interaction:

\bqa
S = \int d^4x \left[\frac12\,\left(\pr_\mu\phi\right)^2 - \frac13 \, \lambda \, \phi^3 \right].
\eqa
For small $\lambda$ this theory describes phonons in crystals with slight non--linearity.

\subsection{Operator derivation}

The general method of the derivation of the KE in the operator formalism can be found in
\cite{Zubarev}. Here we concisely perform the calculations for the particular model in question.
We consider the case when the distribution function is homogeneous in space, i.e. depends only
on time, but not on spacial coordinates.

The normal ordered Hamiltonian for this theory looks as:

\bqa
H = \int d^3 k \, k \, :a_k^+ \, a_k: + H_{int}, \quad k = \left|\vec{k}\right|, \quad \left[a_k, a^+_q\right] = \delta\left(\vec{k} - \vec{q}\right) \nonumber \\
H_{int} = \frac{\lambda}{3} \, \int \frac{d^3 k_1 \, d^3 k_2 \, d^3 k_3}{\sqrt{k_1\, k_2 \, k_3}}\times \nonumber \\ \left[3\, \delta\left(-\vec{k}_1 + \vec{k}_2 + \vec{k}_3\right) \, \left( e^{- i\, \left(-k_1 + k_2 + k_3\right)\, t} \, :a^+_{k_1}\, a_{k_2}\, a_{k_3}: + e^{i\, \left(-k_1 + k_2 + k_3\right)\, t}\,:a_{k_1}\, a^+_{k_2}\, a^+_{k_3}:\right) \right. + \nonumber \\ + \left. \delta\left(\vec{k}_1 + \vec{k}_2 + \vec{k}_3\right)\, \left(e^{- i\, \left(k_1 + k_2 + k_3\right)\, t}\,:a_{k_1}\, a_{k_2}\, a_{k_3}: + e^{i\, \left(k_1 + k_2 + k_3\right)\, t}\,:a^+_{k_1}\, a^+_{k_2}\, a^+_{k_3}:\right)\right].
\eqa
Usually one drops the last two terms in $H_{int}$ due to energy--momentum conservation.
However, we keep them to show the difference of the situation in flat space wrt the curved one.

Consider this theory in some non--stationary initial state, characterized possibly by some density matrix. We would like to find how the distribution function $n_p = \frac{1}{V}\,\left\langle a^+_p\, a_p\right\rangle$ evolves in time. Here $V$ is the volume of space, which appears because $\left\langle a^+_p\, a_q\right\rangle = n_p \, \delta\left(\vec{p} - \vec{q}\right)$ and $V = \delta(0)$.
The evolution equation in the interacting picture is:

\bqa\label{beging}
\frac{dn_p}{dt} = \frac{i}{V}\, \left\langle\left[H_{int}, \, a^+_p \, a_p\right]\right\rangle.
\eqa
Then:

\bqa\label{eqtoclose}
\frac{dn_p}{dt} = \frac{i \, \lambda}{V}\,\int \frac{d^3k_1\, d^3k_2}{\sqrt{k_1\, k_2\, p}}\times \nonumber \\ \left[- \delta\left(-\vec{p} + \vec{k}_1 + \vec{k}_2\right)\, \left(\left\langle :a^+_p \, a_{k_1} \, a_{k_2}: \right\rangle\, e^{- i\, \left(-p + k_1 + k_2\right)\, t} - \left\langle :a_p \, a^+_{k_1} \, a^+_{k_2}: \right\rangle \, e^{i\, \left(-p + k_1 + k_2\right)\, t} \right) +\right. \nonumber \\ +  2\, \delta\left(-\vec{k}_1 + \vec{k}_2 + \vec{p}\right)\, \left(\left\langle :a^+_{k_1} \, a_{k_2} \, a_{p}: \right\rangle\, e^{- i\, \left(-k_1 + k_2 + p\right)\, t} - \left\langle :a_{k_1} \, a^+_{k_2} \, a^+_p: \right\rangle \, e^{i\, \left(-k_1 + k_2 + p\right)\, t}\right) + \nonumber \\ + \left. \delta\left(\vec{p} + \vec{k}_1 + \vec{k}_2\right)\, \left(\left\langle :a_p \, a_{k_1} \, a_{k_2}: \right\rangle\, e^{- i\, \left(p + k_1 + k_2\right)\, t} - \left\langle :a^+_p \, a^+_{k_1} \, a^+_{k_2}: \right\rangle \, e^{i\, \left(p + k_1 + k_2\right)\, t}\right)\right].
\eqa
We dropped the zero modes here.
Careful study of these modes reveals the classical instability of the theory with cubic self--interaction.
However, the terms in the CI which are responsible for this instability are not universal. Such terms in the $\phi^4$ theory do not present, because one always can choose the vacuum value of the scalar field, which respects $Z_2$ symmetry.

The equation (\ref{eqtoclose}) does not close wrt $\left\langle a^+_p\, a_q\right\rangle$. So we have to find $\left\langle a^+_p \, a^+_q \, a^+_k \right\rangle(t)$, $\left\langle a_p \, a^+_q \, a^+_k \right\rangle(t)$ and etc..
Consider e.g.

\bqa\label{20}
\frac{d}{dt} \left\langle :a^+_{k_1} \, a_{k_2} \, a_{k_3}: \right\rangle = i \, \left\langle \left[H_{int}, \, :a^+_{k_1} \, a_{k_2} \, a_{k_3}:\right]\right\rangle
\eqa
After the calculation of the commutator in this expression we see that its RHS depends on such quantities as e.g. $\left\langle a_{k_1} \, a_{k_2} \, a^+_{k_3} \, a^+_{k_4}\right\rangle$. Hence, we have to derive the evolution equations for them and so on. This way one obtains the so called Bogolyubov hierarchy. To truncate the hierarchy one should decide what is the approximation he would like to consider. If one would like to find the CI (the RHS of \eq{beging}) at the order $\lambda^2$ he has to truncate the sequence of these equations already in \eq{20}. This can be done via the application of the Wick's theorem, e.g.:

\bqa
\left\langle a_{k_1} \, a^+_{k_2} \, a_{k_3} \, a^+_{k_4}\right\rangle =  \left\langle a_{k_1}\, a^+_{k_4}\right\rangle\,\left\langle a^+_{k_2}\, a_{k_3}\right\rangle + \left\langle a_{k_1}\, a^+_{k_2}\right\rangle\,\left\langle a_{k_3}\, a^+_{k_4}\right\rangle
\eqa
and so on. The justification of such an anzats/approximation is given in the next subsection of the Appendix (see as well \cite{Mattuck}).

So, if one uses the Wick's theorem in the commutator $\left\langle \left[H_{int}, \, a^+_{k_1} \, a_{k_2} \, a_{k_3} \right]\right\rangle$,
he obtains:

\bqa
\frac{d}{dt} \left\langle :a^+_{k_1} \, a_{k_2} \, a_{k_3}: \right\rangle = - \frac{i\, 2\, \lambda}{\sqrt{k_1\, k_2\, k_3}} \, \delta\left(-\vec{k}_1 + \vec{k}_2 + \vec{k}_3\right) \, e^{i\, \left(-k_1 + k_2 + k_3\right)\, t} \times \nonumber \\ \times \left[(1+n_{k_1})\, n_{k_2}\, n_{k_3}^{\phantom{\frac12}} - n_{k_1}\, (1+n_{k_2})\, (1+n_{k_3})\right],
\eqa
where apart from $\left\langle a^+_p\, a_q\right\rangle = n_p \, \delta\left(\vec{p} - \vec{q}\right)$
we have used that $\left\langle a_p\, a^+_q\right\rangle = (1 + n_p) \, \delta\left(\vec{p} - \vec{q}\right)$. Similarly we can find the equation for $\left\langle a_{k_1} \, a^+_{k_2} \, a^+_{k_3} \right\rangle$. As the result:

\bqa
e^{- i\, \left(-k_1 + k_2 + k_3\right)\, t}\,\left\langle :a^+_{k_1} \, a_{k_2} \, a_{k_3}: \right\rangle(t) - e^{i\, \left(-k_1 + k_2 + k_3\right)\, t}\,\left\langle :a_{k_1} \, a^+_{k_2} \, a^+_{k_3}: \right\rangle(t) = \nonumber \\ - \frac{i\, 2\, \lambda}{\sqrt{k_1\, k_2\, k_3}} \, \delta\left(-\vec{k}_1 + \vec{k}_2 + \vec{k}_3\right) \, {\rm Re} \int_{t_0}^t dt'e^{- i\, \left(- k_1 + k_2 + k_3\right)\, (t-t')}\times \nonumber \\ \times \left[(1+n_{k_1})\, n_{k_2}\, n_{k_3}^{\phantom{\frac12}} - n_{k_1}\, (1+n_{k_2})\, (1+n_{k_3})\right](t').
\eqa
Here $t_0$ is the moment of time when we switch on interactions.

Along the same lines, one can find the answer for $e^{- i\, \left(k_1 + k_2 + k_3\right)\, t}\,\left\langle :a_{k_1} \, a_{k_2} \, a_{k_3}:^{\phantom{+}} \right\rangle - e^{i\, \left(k_1 + k_2 + k_3\right)\, t}\left\langle :a^+_{k_1} \, a^+_{k_2} \, a^+_{k_3}: \right\rangle$. Thus, we obtain the following equation:

\bqa\label{23}
p\,\frac{d n_p}{dt} = - 4\, \lambda^2 \, \int\frac{d^3k_1 \, d^3k_2}{k_1\, k_2} \times \nonumber \\  \left\{\delta\left(-\vec{p} + \vec{k}_1 + \vec{k}_2\right) \, \int_{t_0}^t dt' \cos\left[\left(-p + k_1 + k_2\right)\, (t-t')\right] \, \left[(1+n_p)\, n_{k_1}\, n_{k_2}^{\phantom{\frac12}} - n_p \, (1+n_{k_1})\, (1+n_{k_2})\right](t') \right. \nonumber \\ + 2\, \delta\left(-\vec{k}_1 + \vec{k}_2 + \vec{p}\right)\,\int_{t_0}^t dt' \cos\left[\left(- k_1 + k_2 + p\right)\, (t-t')\right] \, \left[n_{k_1}\,(1 + n_{k_2})\, (1 + n_p)^{\phantom{\frac12}} - (1+n_{k_1})\, n_{k_2} \, n_p\right](t') \nonumber \\ +
\left. \delta\left(\vec{k}_1 + \vec{k}_2 + \vec{p}\right)\,\int_{t_0}^t dt' \cos\left[\left(k_1 + k_2 + p\right)\, (t-t') \right] \, \left[(1 + n_{k_1})\,(1 + n_{k_2})\, (1 + n_p)^{\phantom{\frac12}} - n_{k_1}\, n_{k_2} \, n_p\right](t') \right\}.
\eqa
We assume that $n_k$'s depend on $t$ slowly and can be taken out from the integral over $t'$. Then we take $t\to+\infty$ and $t_0\to -\infty$. The result of the integration over $t'$ is the minus $\delta$--function ensuring the energy conservation in the corresponding process. Hence, the last term in \eq{23} vanishes because it describes processes forbidden by the energy--momentum conservation. In the case of massless theory the first two terms describe allowed processes and the KE acquires the form:

\bqa\label{24}
p\,\frac{d n_p}{dt} = 4\, \lambda^2 \, \int\frac{d^3k_1 \, d^3k_2}{k_1\, k_2} \times \nonumber \\  \left\{\delta\left(-\vec{p} + \vec{k}_1 + \vec{k}_2\right) \, \delta\left(-p + k_1 + k_2\right) \, \left[(1+n_p)\, n_{k_1}\, n_{k_2}^{\phantom{\frac12}} - n_p \, (1+n_{k_1})\, (1+n_{k_2})\right] \right. \nonumber \\ + \left. 2\, \delta\left(-\vec{k}_1 + \vec{k}_2 + \vec{p}\right)\,\delta\left(- k_1 + k_2 + p\right)\, \left[n_{k_1}\,(1 + n_{k_2})\, (1 + n_p)^{\phantom{\frac12}} - (1+n_{k_1})\, n_{k_2} \, n_p\right]  \right\}.
\eqa
Having understood the above equation and the physical meaning of all terms in it (see the discussion in the main body of the text) it is not hard to write the KE for the massive real scalar $\phi^4$ theory:

\bqa\label{25}
\omega\,\frac{d n_p}{dt} = 4\, \lambda^2 \, \int\frac{d^3k_1 \, d^3k_2\, d^3 k}{\omega_1\, \omega_2\, \omega_3} \times \nonumber \\  \left\{3\,\delta\left(-\vec{p} - \vec{k}_1 + \vec{k}_2 + \vec{k}_3\right) \, \frac{\sin\left[\left(- \omega - \omega_1 + \omega_2 + \omega_3\right)\, (t-t_0)\right]}{- \omega - \omega_1 + \omega_2 + \omega_3} \times \right.  \nonumber \\ \times \left[(1+n_\omega)\, (1 + n_{\omega_1})\, n_{\omega_2}\, n_{\omega_3}^{\phantom{\frac12}} - n_\omega \, n_{\omega_1} \, (1+n_{\omega_2})\, (1+n_{\omega_3})\right] \nonumber \\ + 3\, \delta\left(-\vec{k}_1 + \vec{k}_2 + \vec{k}_3 + \vec{p}\right)\, \frac{\sin\left[\left(- \omega_1 + \omega_2 + \omega_3 + \omega\right)\, (t-t_0)\right]}{- \omega_1 + \omega_2 + \omega_3 + \omega} \times \nonumber \\ \times \left[n_{\omega_1}\,(1 + n_{\omega_2})\, (1 + n_{\omega_3}) (1 + n_\omega)^{\phantom{\frac12}} - (1+n_{\omega_1})\, n_{\omega_2} \, n_{\omega_3}\, n_\omega\right]
\nonumber \\ + \delta\left(-\vec{p} + \vec{k}_1 + \vec{k}_2 + \vec{k}_3\right)\, \frac{\sin\left[\left(- \omega + \omega_1 + \omega_2 + \omega_3\right)\, (t-t_0)\right]}{- \omega + \omega_1 + \omega_2 + \omega_3} \times \nonumber \\ \times \left[(1+n_{\omega})\, n_{\omega_1} \, n_{\omega_2}\, n_{\omega_3} - n_{\omega}\,(1 + n_{\omega_1})\, (1 + n_{\omega_2}) (1 + n_{\omega_3})^{\phantom{\frac12}}\right]
\nonumber \\ +
\delta\left(\vec{k}_1 + \vec{k}_2 + \vec{k}_3 + \vec{p}\right)\,\frac{\sin\left[\left(\omega_1 + \omega_2 + \omega_3 + \omega \right)\, (t-t_0) \right]}{\omega_1 + \omega_2 + \omega_3 + \omega} \times \nonumber \\ \left. \times \left[(1 + n_{\omega_1})\,(1 + n_{\omega_2})\, (1 + n_{\omega_3}) \, (1 + n_{\omega})^{\phantom{\frac12}} - n_{\omega_1}\, n_{\omega_2} \, n_{\omega_3} \, n_\omega\right] \right\},
\eqa
where $\omega^2 = \vec{p}^2 + m^2$. As $t-t_0 \to \infty$ the function $\frac{\sin \Delta \omega \, (t-to)}{\Delta \omega}$ is reduced to the $\delta$--functions ensuring energy conservation. The only allowed by the energy--momentum conservation process is the scattering between the scalar particles, i.e. the CI contains only $\delta\left(\omega_1 + \omega_2 - \omega_3 - \omega\right) \, \left[n_{\omega_1}\, n_{\omega_2} \, (1 + n_{\omega_3})\, (1 + n_\omega) - (1 + n_{\omega_1})\, (1 + n_{\omega_2})\, n_{\omega_3}\, n_\omega\right]$. Two more terms will stay in the CI for the massless $\phi^4$ theory: only the last term in the above CI is forbidden by the energy--momentum conservation in the massless theory.

In the small density limit, $n_\omega\ll 1$, for all $\omega$ we can neglect $n_\omega$ in comparison with 1 in all expressions inside the CI\footnote{This is, in fact, proper classical limit, because otherwise due to high particle density quantum coherence between particles is not destroyed.}.
Then for the massive $\phi^4$ theory the CI will reduce to its appropriate classical Boltzmann form $\delta\left(\omega_1 + \omega_2 - \omega_3 - \omega\right) \, \left[n_{\omega_1}\, n_{\omega_2} -  n_{\omega_3}\, n_\omega\right]$. Now one can immediately see that because of the energy conservation  this expression vanishes for the equilibrium Boltzmann
distribution $n_\omega \propto e^{-\frac{\omega}{T}}$, for some constant $T$. As well it is not hard to see that all terms in the CI \eq{24} and \eq{25} (for $(t-t_0)\to \infty$) vanish when $n_\omega = \frac{1}{e^{\omega/T} - 1}$.

Finally, it is not hard to generalize the KE to the spatially inhomogeneous case by the calculation of $\pr_\mu n = \frac{1}{V}\, \left\langle \left[P_\mu, a^+\, a\right]\right\rangle$, where $P_\mu = \left(H,P_i\right)$ is the momentum operator. Then the LHS of \eq{24} and \eq{25} will change to $p^\mu \, \pr_\mu n_p(x)$. The RHS will be the same.

\subsection{Derivation from the Dyson-Schwinger equation}

The general method to derive the KE from the DSE of the in--in (non--stationary or Schwinger--Keldysh) formalism can be found in e.g. \cite{LL} or \cite{Kamenev}. Here we concisely repeat that derivation. The main reason to present this derivation here is to show the relation of the KE to the summation of the loop diagrams in the IR limit. The technical explanation of the relation between the Wick contractions, which were used in the derivation of the Boltzmann equation, and the partial resummation of loop graphs can be found in \cite{Mattuck}.

The reason for the Schwinger--Keldysh diagrammatic technic can be seen from the following observation. Suppose we would like to calculate the expectation value of an operator ${\cal O}$ at some moment of time $t$ wrt some state $\Psi$

\bqa
\left\langle {\cal O} \right\rangle(t) \equiv \left\langle \Psi\left| \overline{T} e^{i\,\int_{t_0}^t dt' H(t')}\,{\cal O} \, T e^{-i\,\int_{t_0}^t dt' H(t')}\right| \Psi\right\rangle,
\eqa
where $H(t)$ is the full Hamiltonian of the system, which may depend on time due to the presence of e.g. time dependent background fields, $\overline{T}$ is anti--time ordering, $t_0$ is initial moment of time when we know the expectation value. Transferring to the interaction picture, we get:

\bqa
\left\langle {\cal O} \right\rangle(t) = \left\langle \Psi\left| S^+(t, t_0)\, T\left[ {\cal O}_0(t) \, S(t,t_0)\right]\right| \Psi\right\rangle = \nonumber \\ = \left\langle \Psi\left| S^+(t, t_0) S^+(+\infty, t)\, S(+\infty, t)\, T\left[ {\cal O}_0(t) \, S(t,t_0)\right]\right| \Psi\right\rangle = \nonumber \\ = \left\langle \Psi\left| S^+(+\infty, t_0)\, T\left[ {\cal O}_0(t) \, S(+\infty,t_0)\right]\right| \Psi\right\rangle,
\eqa
where $S(t,t_0) = T e^{-i\,\int_{t_0}^t dt' H_{int}(t')}$ and ${\cal O}_0(t)$ is the time dependence of the operator ${\cal O}$ in the interaction picture. Adiabatic switching off of the interactions is assumed in the future infinity.

If we adiabatically switch on the interaction around the time $t_0$, then we can write the expectation value as:

\bqa\label{calo}
\left\langle {\cal O} \right\rangle(t) = \left\langle \Psi\left| S^+(+\infty, -\infty)\, T\left[ {\cal O}_0(t) \, S(+\infty,-\infty)\right]\right| \Psi\right\rangle.
\eqa
Good question is that if one can take $t_0$ to $-\infty$. If the state of the theory does become stationary (e.g. thermalizes), then $t_0$ can be taken to the past infinity. However, if the state does not get stationary, which may be the case if the background field is never switched off, then $t_0$ can not be taken to minus infinity because of the explosive behavior of the correlation functions when $(t-t_0)\to \infty$.

Now if $\left|\Psi\right\rangle$ were the true vacuum state $\left|vacuum \right\rangle$ of the free theory, then by adiabatic switching on and off the interactions one could not disturb such a state, i.e. $\left\langle vacuum \left| S^+(+\infty, -\infty)\right| excited \,\, state\right\rangle = 0$, while $\left|\left\langle vacuum \left| S^+(+\infty, -\infty)\right| vacuum \right\rangle\right| = 1$ and

\bqa
\left\langle {\cal O} \right\rangle(t) = \sum_{state} \left\langle vacuum \left| S^+(+\infty, -\infty)\right| state \right\rangle \, \left\langle state \left| T\left[ {\cal O}_0(t) \, S(+\infty,-\infty)\right]\right| vacuum \right\rangle = \nonumber \\ = \left\langle vacuum \left| S^+(+\infty, -\infty)\right| vacuum \right\rangle \, \left\langle vacuum \left| T\left[ {\cal O}_0(t) \, S(+\infty,-\infty)\right]\right| vacuum \right\rangle = \nonumber \\ =
\frac{\left\langle vacuum \left| T\left[ {\cal O}_0(t) \, S(+\infty,-\infty)\right]\right| vacuum \right\rangle}{\left\langle vacuum \left| S(+\infty, -\infty)\right| vacuum \right\rangle} .
\eqa
This way we arrive at the standard Feynman diagrammatic technic, which is based only on the T--ordered quantities.

However, if $\left|\Psi\right\rangle$ is not a stable state one can not use the above machinery and has
to deal directly with \eq{calo}. One has to expand both $S$ and $S^+$ in powers of the interaction Hamiltonian and apply the Wick's theorem. Then one will encounter two types of vertices and four types of propagators. The vertices will be coming from $S$ and $S^+$. At the same time propagators appear from the Wick contractions inside $S$ (time ordered), from those inside $S^+$ (anti--time ordered) and from Wick contractions between $S$ and $S^+$. However, there are only three independent propagators.

The functional integral in such a situation has the action, which schematically can be written as \cite{Kamenev}:

\bqa
S = \int_{C} dt \int d^3x \left[\frac12 \, \left(\pr_\mu \phi\right)^2 - \frac{\lambda}{3}\, \phi^3\right],
\eqa
where $C$ is the Keldysh time contour running from $-\infty$ to $+\infty$ and back. That is due to the presence of both $S$ and $S^+$ in \eq{calo}. The exact expression for the action will be given in a moment.

This action can be rewritten as:

\bqa
S = \int_{-\infty}^{+\infty} dt \int d^3x \left[\frac12 \, \left(\pr_\mu \phi_{+}\right)^2 - \frac12 \, \left(\pr_\mu \phi_{-}\right)^2 - \frac{\lambda}{3}\, \phi_{+}^3 + \frac{\lambda}{3}\, \phi_{-}^3\right],
\eqa
where $\phi_{+}$ is the field on the direct part of the time contour, while $\phi_{-}$ is the field on the reverse part of it. The kinetic term in this equation again is written schematically \cite{Kamenev}.
After the Keldysh rotation of the fields:

\bqa
\phi_{cl} = \frac{1}{\sqrt{2}}\, \left[\phi_{+} + \phi_{-}\right], \quad \phi_q = \frac{1}{\sqrt{2}}\, \left[\phi_{+} - \phi_{-}\right]
\eqa
the precise form of the action is as follows \cite{Kamenev}:

\bqa
S = \int d^4x \int d^4y \left(\phi_{cl}(x)^{\phantom{\frac12}}, \phi_q(x)\right)\, \left(
  \begin{array}{cc}
    0 & \left[D_0^A\right]^{-1}(x,y) \\
    \left[D_0^R\right]^{-1}(x,y) & \left[D_0^{-1}\right]^K(x,y) \\
  \end{array}
\right) \, \left(
  \begin{array}{c}
    \phi_{cl}(y)  \\
    \phi_q(y) \\
  \end{array}
\right) - \nonumber \\ - 2\, \lambda \, \int d^4 x \left[\phi_{cl}^2(x) \, \phi_q (x) + \frac13 \, \phi_q^3(x)\right].
\eqa
The vertices and propagators in this theory are shown in the figure.
Here $D_0^A$ and $D_0^R$ are the advanced and retarded propagators, whose Fourier transforms look as

\begin{figure}[t]
\begin{center} \includegraphics[width=300pt]{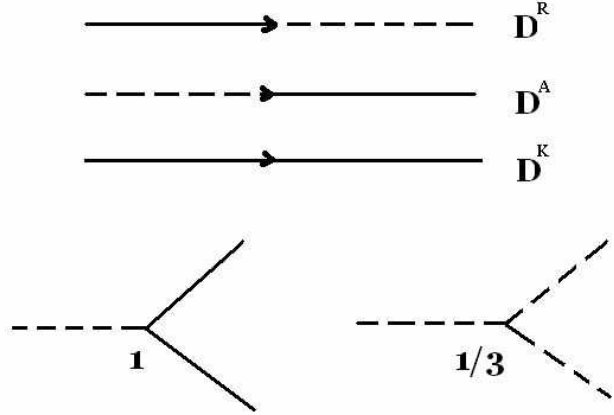}
\caption{The solid line corresponds to $\phi_{cl}$, while the dashed line --- to $\phi_q$.}
\end{center}
\end{figure}

\bqa
D_0^{R,A}(\omega, k) = \frac{1}{\left(\epsilon \pm i\, 0\right)^2 - \vec{k}^2}.
\eqa
In the $x$--space they have the form

\bqa
\left[D_0^{R,A}(x,y)\right]^{-1} = \theta(\pm \Delta t) \, \delta^{(4)}(x-y) \, \Box,
\eqa
where $\Box$ is the D'Alamber's operator. The retarded and advanced propagators
carry information about the spectrum
of the theory and have the following relevant for us properties:

\bqa
D^A_0(x,y)\, D^R_0(x,y) = 0, \quad D^R_0(t,t) + D^A_0(t,t) = 0, \quad \left[D^A_0\right]^+ = D_0^R,
\eqa
which remain valid even for their quantum corrected versions \cite{Kamenev}.

The Keldysh propagator $\left[D^K_0\right]^+ = - D^K_0$ carries statistical information about the theory: it shows which levels from the spectrum are occupied. By definition $D^K_0(\omega, k) \equiv \int d^4x \, e^{i\, k_\mu \, x^\mu}\, \left\langle \left\{\phi(x), \phi(0)\right\}\right\rangle$,
where $\{\cdot, \cdot\}$ means the anti--commutator.
For the thermal state it acquires the form:

\bqa
D^K_0(\omega, k) = \coth\frac{\omega}{2\, T}\, \left[D^R_0(\omega,k) - D^A_0(\omega,k)\right] = \coth\frac{\omega}{2\, T}\,\delta\left(\omega^2 - \vec{k}^2\right),
\eqa
This propagator is present in the above action to regularize the functional integral \cite{Kamenev}.

The last expression allows to guess the ansatz for the full quantum Keldysh propagator \cite{Kamenev}, when one is close to the stationary situation:

\bqa
D^K(x,y) = \int d^4 z\left[D_0^R(x,z)\,F(z,y) - F(x,z)\, D_0^A(z,y)\right] \equiv \left[D_0^R \circ F - F\circ D_0^A\right](x,y)
\eqa
with some unknown in general kernel $F(x,y)$ which characterizes the statistical distribution.
In the stationary situation the Fourier transform of $F(x,y)$ is $\coth\frac{\omega}{2\, T} = 1 + 2\, n_\omega$, where $n_\omega$ is the Plankian distribution function.
Furthermore,

\bqa
\left[D_0^{-1}\right]^K = - \left[D_0^R\right]^{-1}\circ D_0^K \circ \left[D_0^A\right]^{-1} = \left[D_0^R\right]^{-1} \circ F_0 - F_0\circ \left[D_0^A\right]^{-1}.
\eqa
One can write the DSE for the full quantum matrix propagator:

\bqa
\hat{D} \equiv \left(
  \begin{array}{cc}
    D^K & D^R \\
    D^A & 0 \\
  \end{array}
\right)
\eqa
which, as can be shown \cite{Kamenev}, keeps the same form as $\hat{D}_0$. The equation looks as:

\bqa
\left(\hat{D}_0^{-1} - \hat{\Sigma}\right)\circ \hat{D} = 1,
\eqa
where

\bqa
\hat{\Sigma} = \left(
  \begin{array}{cc}
    0 & \Sigma^A \\
    \Sigma^R & \Sigma^K \\
  \end{array}
\right)
\eqa
is the self--energy matrix. It can be shown that it has the same form and properties as $\hat{D}_0^{-1}$. Below we will see this fact at one loop level. In components of $\hat{D}$ and $\hat{\Sigma}$ the DSE can be written as

\bqa
\Box D^{R,A} - \Sigma^{R,A}\circ D^{R,A} = 1, \nonumber \\
\left[\Box,F\right] = \Sigma^K - \left(\Sigma^R \circ F - F\circ \Sigma^A \right)
\eqa
if one uses the ansatz $D^K = D^R\circ F - F\circ D^A$ and neglects $\left[D^{-1}_0\right]^K$ in comparison with $\Sigma^K$, because it is just a regulator.

Let us derive explicitly the DSE in one loop approximation.
Note that while considering full $D^K$ we take the bare $D^{A,R}_0$. The justification of this approximation comes from the observation
that the only way renormalization appears in $D^A$ and $D^R$ is through the change of the spectrum --- renormalization of the mass and etc.. So once we know the spectrum of quasiparticles precisely it means that we know the retarded and advanced propagators as classical objects. As well we substitute into the DSE the IR, i.e. renormalized, value of the vertex in the theory.  In such circumstances the DSE becomes an equation for $D^K$ only --- for the propagator containing statistical information about the state in the theory.

First we calculate the self--energy at one
loop order:

\begin{itemize}

\item It is straightforward to see that $\Sigma_{cl-cl}$ is zero (see fig.):

\bqa
\Sigma_{cl-cl} \propto \lambda^2 \, D^R_0(x,y)\, D^A_0(x,y) = 0,
\eqa
because of the presented above properties of $D_0^R$ and $D_0^A$.

\begin{figure}[t]
\begin{center} \includegraphics[width=300pt]{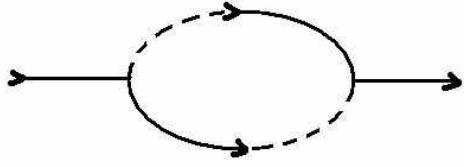}
\caption{The graph defining $\Sigma_{cl-cl}$.}
\end{center}
\end{figure}

\item At the same time (see fig.)

\bqa
\Sigma_{cl-q} \equiv \Sigma^A(x,y) = 4\, i\, \lambda^2 \, D^A_0(x,y)\, D^K(x,y) \neq 0
\eqa
Since $\Sigma^A \sim D^A \sim \theta(t_y - t_x)$, this quantity is indeed of an advanced type and should, as it does, stand in the upper triangular corner of the $\hat{\Sigma}$ matrix.
The prefactor 4 comes from 4 ways of choosing external legs, 2 internal permutations and $1/2!$ for having two identical vertexes.

\begin{figure}[t]
\begin{center} \includegraphics[width=300pt]{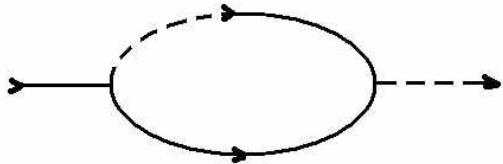}
\caption{The graph defining $\Sigma^A$.}
\end{center}
\end{figure}

\item Similarly (see fig.)

\bqa
\Sigma_{q-cl} \equiv \Sigma^R(x,y) = 4\, i\, \lambda^2 \, D^R_0(x,y)\, D^K(x,y)
\eqa
As well $\Sigma^R \sim D^R \sim \theta(t_x - t_y)$. Hence, at the one loop level $\Sigma^R = \left[\Sigma^A\right]^+$.

\begin{figure}[t]
\begin{center} \includegraphics[width=300pt]{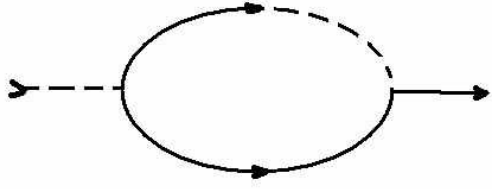}
\caption{The graph defining $\Sigma^R$.}
\end{center}
\end{figure}

\item And finally (see fig.):

\bqa
\Sigma_{q-q} \equiv \Sigma^K(x,y) = 2\, i\, \lambda^2 \left[D^K(x,y)\right]^2 + 6\, i\, \frac{\lambda}{3}\, \lambda \, \left[D^A(x,y)\right]^2 + 6\, i\, \lambda\, \frac{\lambda}{3}\, \left[D^R(x,y)\right]^2 = \nonumber \\ = 2\, i\, \lambda^2 \left(\left[D^K(x,y)\right]^2 + \left[D^R(x,y) - D^A(x,y)\right]^2\right),
\eqa
where at the last step we have used the property $D^R(x,y)\, D^A(x,y) = 0$. Now one can see that because $\left[D^R\right]^+ = D^A$ and $\left[D^K\right]^+ = - D^K$ we have that $\left[\Sigma^K\right]^+ = - \Sigma^K$.

\begin{figure}[t]
\begin{center} \includegraphics[width=300pt]{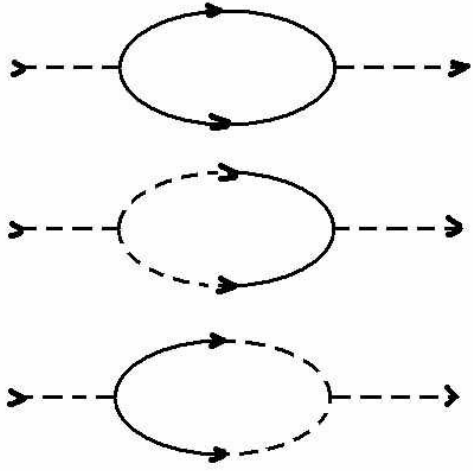}
\caption{The graphs defining $\Sigma^K$.}
\end{center}
\end{figure}

\end{itemize}

Thus, as promised at one loop level $\hat{\Sigma}$ has the same properties as $\hat{D}_0^{-1}$.
Now we plague these expressions for the components of $\hat{\Sigma}$ into the DSE for $F$ and use the Wigner transformation:

\bqa
A\left(x,y\right) = A\left(X, \chi\right), \quad X=\frac{x+y}{2},\quad \chi=x-y \nonumber \\
A\left(X,\chi\right) \equiv \int d^4k \,e^{i\, k^\mu \chi_\mu}\, a\left(X, k\right).
\eqa
Here $a$ is the Wigner transform of $A$.

The Wigner transformation has the following properties \cite{Kamenev}. If

\bqa
A\left(X,\chi\right) = \int d^4k \, e^{i\, k^\mu \chi_\mu}\, a\left(X, k\right), \quad B\left(X,\chi\right) = \int d^4k \, e^{i\, k^\mu \chi_\mu}\, b\left(X, k\right),
\eqa
and if $C(x,y) = A(x,y)\, B(x,y)$, then

\bqa
c\left(X,k\right) \equiv \int d^4\chi\, e^{-i\, k^\mu\, \chi_\mu} \, C(X,\chi) = \int d^4k_1\, d^4k_2 \, \delta^{(4)}(k-k_1-k_2)\,a\left(X,k_1\right) \, b\left(X,k_2\right).
\eqa
At the same time if $C = A\circ B$, then:

\bqa
c\left(X,k\right) = a\left(X,k\right)\, e^{-i\, \left(\overleftarrow{\pr_X} \, \overrightarrow{\pr_k} - \overleftarrow{\pr_k} \, \overrightarrow{\pr_X}\right)}\, b\left(X,k\right) \approx  a\left(X,k\right)\, b\left(X,k\right) + \dots
\eqa
The last approximation is valid if $a$ and $b$ are relatively slow functions of $X=(x+y)/2$ and fast functions of $\chi = x-y$. Thus, we should have a separation of scales in the problem under consideration.

Using this approximation, the derived above expressions for the self--energy and defining as $d^A$, $d^R$ and $f$ the Wigner transforms of $D^A_0$, $D^R_0$ and $F$, correspondingly, we obtain that:

\bqa
\Sigma^R\circ F - F\circ \Sigma^A \rightarrow 4\, i\,\lambda^2\, f\left(X,k\right)\,
\int d^4k_1\, d^4k_2 \, \delta^{(4)}\left(k - k_1 - k_2\right)\,\Delta_d\left(X,k_1\right) \,
d^K\left(X,k_2\right) \approx \nonumber \\
2\, i\,\lambda^2\, f\left(X,k\right) \,\int d^4k_1\, d^4k_2 \,\delta^{(4)}\left(k - k_1 - k_2\right)\,\Delta_d\left(X,k_1\right)\,\Delta_d\left(X,k_2\right)\, \left[f\left(X,k_2\right) + f\left(X,k_1\right)\right]
\eqa
where $\Delta_d = \left[d^R - d^A\right]$ and on the last step we have substituted the expression for $d^K$ through $d^R$, $d^A$ and $f$. At the end we performed the symmetrization of the argument of $f$ under exchange of $k_1\leftrightarrow k_2$.

Similarly:

\bqa
\Sigma^K \rightarrow 2\, i\, \lambda^2\, \int d^4k_1\, d^4k_2\, \delta^{(4)}\left(k - k_1 - k_2\right)\,\Delta_d\left(X;k_1\right)\, \Delta_d\left(X;k_2\right)\,\left[f\left(X;k_1\right)\, f\left(X;k_2\right) + 1\right].
\eqa
And finally,

\bqa
\left[F, \, \Box\right] \rightarrow - 2\, i\, k^\mu\, \frac{\pr}{\pr X^\mu} \, f\left(X;k\right).
\eqa
Putting all this together, we obtain:

\bqa\label{77}
k^\mu \, \pr_\mu \, f\left(X,k\right) = 2\, \lambda^2 \, \int d^4k_1\, d^4k_2 \, \delta^{(4)}\left(k-k_1-k_2\right)\,\Delta_d(X,k_1)\, \Delta_d(X,k_2)\times \nonumber \\ \times \left\{f\left(X,k_1\right)\, f\left(X,k_2\right)^{\phantom{\frac12}} + 1 - f\left(X,k\right)\, \left[f\left(X,k_1\right) + f\left(X,k_2\right)\right]\right\}
\eqa
Now due to the properties of the propagator $D^K$, the Wigner transform of $F$ obeys:

\bqa
f\left(X,k\right) = - f\left(X,-k\right)
\eqa
Furthermore because $\Delta_d(k) \sim D^R_0(k) - D_0^A(k) \sim \delta\left(k_0^2 - \vec{k}^2\right)$ the $k_1$ and $k_2$ legs are on mass--shell. We as well put $k$ on mass--shell --- $k_0 = \pm \left|\vec{k}\right|$. Then the mass--shell distribution function obeys:

\bqa
f\left(X,\vec{k}\right) = sign(k_0)\, f\left(X,sign(k_0)\, \vec{k}\right).
\eqa
Representing the $f$ function through the distribution function, $f\left(X;k\right) = 1 + 2\, n\left(X;k\right)$, we see that equation \eq{77} reduces to the spatially inhomogeneous form of the  KE obtained in the previous section of the Appendix.

Note that the whole CI comes from imaginary contribution to $\hat{\Sigma}$ \cite{Kamenev}.
Apart from that, the positive contribution to the CI (the gain processes) comes from $\Sigma^K$, while the negative (the loss) one comes from $\Sigma^R\circ F - F\circ \Sigma^A$.

\end{document}